\pdfoutput=1
\documentclass[aps,prd,amsmath,floats,floatfix, twocolumn,
superscriptaddress,nofootinbib,showpacs]{revtex4-1}

\usepackage[T1]{fontenc}
\usepackage[utf8]{inputenc}
\usepackage{lmodern}
\usepackage{verbatim}
\usepackage{widetable}

\usepackage[dvipsnames, usenames]{xcolor}
\definecolor{linkcolor}{rgb}{0.0,0.3,0.5}
\usepackage[hypertexnames=false, unicode, colorlinks=true, linkcolor=linkcolor,
citecolor=linkcolor, filecolor=linkcolor,urlcolor=linkcolor,
pdfusetitle]{hyperref}
\usepackage{orcidlink}

\usepackage[all]{hypcap}
\usepackage{graphicx}
\usepackage{xspace}
\usepackage{amssymb}
\usepackage[normalem]{ulem} 
\usepackage{bm} 
\usepackage{enumitem,amssymb}

\usepackage{microtype}

\usepackage[english]{babel}
\usepackage{blindtext}

\graphicspath{%
  {figs/}%
}

\DeclareMathAlphabet{\mathpzc}{OT1}{pzc}{m}{it}

\newlist{todolist}{itemize}{2}
\setlist[todolist]{label=$\square$}
\newcommand{\ssec}[1]{\section{#1}}
\newcommand{\heavyBBHscalarlimits}{{$[0.55, 3.3]\times 10^{-13}$}}
\newcommand{\lightBBHscalarlimits}{{$[2.2, 11]\times 10^{-13}$}}
\newcommand{\heavyBBHvectorlimits}{{$[0.11, 4.9]\times 10^{-13}$}}
\newcommand{\lightBBHvectorlimits}{{$[0.43, 18]\times 10^{-13}$}}

\newcommand{\mb}{m_{b}}

\newcommand{\scalarlimits}{{$[0.55, 11]\times 10^{-13}$}}
\newcommand{\vectorlimits}{{$[0.11, 18]\times 10^{-13}$}}

\begin{document}

\title{Ultralight boson constraints from gravitational wave observations \\of spinning binary black holes}

\author{P.S. Aswathi\orcidlink{0009-0008-1458-3338}}
\affiliation{OzGrav-ANU, Centre for Gravitational Astrophysics, College of Science, The Australian National University, Australian Capital Territory 2601, Australia}

\author{William E. East\orcidlink{0000-0002-9017-6215}}
\affiliation{Perimeter Institute for Theoretical Physics, Waterloo, Ontario N2L 2Y5, Canada}

\author{Nils Siemonsen\orcidlink{0000-0001-5664-3521}}
\affiliation{Princeton Gravity Initiative, Princeton University, Princeton, New Jersey 08544, USA}
\affiliation{Department of Physics, Princeton University, Princeton, New Jersey 08544, USA}

\author{Ling Sun\orcidlink{0000-0001-7959-892X}}
\affiliation{OzGrav-ANU, Centre for Gravitational Astrophysics, College of Science, The Australian National University, Australian Capital Territory 2601, Australia}

\author{Dana Jones\orcidlink{0000-0003-3987-068X}}
\affiliation{OzGrav-ANU, Centre for Gravitational Astrophysics, College of Science, The Australian National University, Australian Capital Territory 2601, Australia}


\date{\today}
\begin{abstract}
In the presence of an ultralight scalar or vector boson, a spinning black hole will be spun down through the superradiant instability. We use spin measurements from gravitational wave observations of binary black holes, in particular the heavy binary black hole merger event GW231123, along with the lower-mass GW190517 event, to constrain the existence of ultralight bosons. We disfavor scalars with masses in the range of \scalarlimits~eV and vectors in the range of \vectorlimits~eV, making only a conservative assumption that the black hole lifetimes are greater than $10^5$ years. The lower ends of these ranges, where the exclusion confidence is the highest, were not previously excluded by spin measurements from electromagnetic or gravitational wave observations. We map these constraints to axion and dark photon models with interactions.
\end{abstract}
\maketitle

\ssec{Introduction}%
Ultralight bosons arise in a number of extensions of the Standard Model of particle physics. 
The axion offers a solution to the strong \textit{CP} problem in quantum chromodynamics~\cite{Peccei:1977hh, Weinberg:1977ma,Peccei:1977ur}, while a hidden sector of light scalar and vector particles naturally emerges in string theory compactifications~\cite{Arvanitaki:2009fg}. Axionlike particles and dark photons are also considered promising candidates for dark matter~\cite{Goodsell:2009xc,Jaeckel:2010ni,Essig:2013lka,Hui:2016ltb}. However, detecting ultralight bosons with terrestrial experiments is challenging due to their typically feeble couplings to Standard Model fields.

Rotating black holes (BHs) provide a natural setting to probe ultralight bosons through superradiance. If the boson's Compton wavelength is comparable to the BH radius, a superradiant instability can amplify the field, forming a gravitationally bound cloud that extracts angular momentum from the BH~\cite{Brito:2015rjv,Starobinsky:1973aij,1971JETPL..14..180Z,PhysRevLett.28.994,Detweiler:1980uk,Press:1972zz,Bekenstein:1973mi,Arvanitaki:2009fg,Arvanitaki:2010sy,Furuhashi:2004jk,Dolan:2007mj}. The process continues until the BH spin drops to a critical value determined by the boson mass~\cite{Arvanitaki:2009fg, Arvanitaki:2010sy,Brito:2014wla,East:2017ovw,East:2018glu}. The spin-down induced by superradiance can thus serve as an astrophysical probe of ultralight bosons. An absence of rapidly spinning BHs within a certain mass range could signal the existence of a boson whose Compton wavelength matches the BH size, triggering the instability. Conversely, the observation of rapidly spinning BHs in this range can be used to rule out the possibility of bosons with corresponding masses. 

Electromagnetic observations of accreting stellar mass BHs can be used to measure their spins, though there are some uncertainties associated with the unknown details of the accretion and emission physics. These observations disfavor the scalar and vector bosons in the mass range of
$\sim 10^{-13}$--$10^{-11}$~eV, with the caveats associated with such spin measurements~\cite{Arvanitaki:2014wva, Stott:2018opm, Baryakhtar:2020gao, Hoof:2024quk, Baryakhtar:2017ngi, Cardoso:2018tly}.
Here, we focus on using gravitational wave (GW) observations of BH mergers as a way to measure BH spins and place constraints~\cite{Arvanitaki2017}. A previous analysis applied to LIGO and Virgo GWTC-2 observations strongly disfavors scalar bosons in the range $1.3\times10^{-13}$--$2.7\times 10^{-13}$~eV assuming an inspiral timescale of $10^7$ years~\cite{Ng:2020ruv,Ng:2019jsx}. Select events were also used to constrain vector bosons in Ref.~\cite{Ghosh:2021zuf}.

The quasimonochromatic GWs sourced by the time-dependent quadrupole of the boson clouds formed around stellar-mass BHs would also fall within the sensitive band of ground-based detectors~\cite{Arvanitaki:2014wva,Brito:2017zvb,Isi2019,Jones2023}. Blind and targeted searches have been conducted for continuous or long-transient GWs, as well as for the stochastic GW background, potentially generated by scalar and vector clouds. 
These searches lead to constraints on the existence of scalars and vectors, disfavoring a mass range of $\sim 10^{-13}$--$10^{-12}$ eV~\cite{Tsukada:2018mbp, Tsukada:2020lgt, O3_all-sky_scalar_bosons, Palomba2019, Dergachev2019,O3_CWs_Milky_Way, Zhu2020, O2_CygnusX1_scalar_bosons, Collaviti2024}, under certain somewhat optimistic assumptions about the unknown population statistics of BH mass, spin, and age. 

Due to differences in assumptions and analysis techniques, some care is required in comparing different superradiance exclusions across the literature. In Appendix~\ref{app:work_comp}, we provide a more detailed summary of the relevant constraints from prior studies.

In this work, we present constraints directly derived from binary BH mergers. In particular, we analyze two GW events observed during the third and fourth observing runs of Advanced LIGO~\cite{LIGOScientific:2014pky}, Virgo~\cite{VIRGO:2014yos}, and KAGRA~\cite{KAGRA:2020tym} with high measured spins for the constituent BHs.
Combining these two events, we are able to exclude scalars and vectors in a mass range of \scalarlimits~eV and \vectorlimits~eV, respectively, at 90\% confidence. 
These new constraints assume only that the observed GW signals correspond to binary BHs that can be adequately described by the present waveform models, and that the constituent BHs were born (and not significantly spun up, e.g., through accretion) more than $10^5$ years before merging. 


\ssec{GW231123 and GW190517}%
A high-mass binary BH merger, GW231123, was observed by the two Advanced LIGO detectors on November 23, 2023, during the LIGO-Virgo-KAGRA's fourth joint observing run (O4)~\cite{GW231123}. The two constituent BHs have source-frame masses of $M_1 = 137^{+22}_{-17}\ M_\odot$ and $M_2 = 103^{+20}_{-52}\ M_\odot$, and dimensionless spins of $\chi_1 = 0.9^{+0.10}_{-0.19}$ and $\chi_2 = 0.8^{+0.20}_{-0.51}$ (90\% credible intervals)~\cite{GW231123}.
The posteriors of the parameters are obtained by equally weighting and combining results from five waveform models~\cite{GW231123,zenodo_GW231123}. 
During the previous observing run (O3), the two Advanced LIGO detectors and Advanced Virgo observed a lower-mass binary BH merger, GW190517, with $M_1 = 39.2^{+13.9}_{-9.2}\ M_\odot$, $M_2 = 24.0^{+7.4}_{-7.9}\ M_\odot$, $\chi_1 = 0.9^{+0.09}_{-0.30}$, and $\chi_2 = 0.62^{+0.34}_{-0.54}$ (90\% credible intervals), calculated from posterior samples drawn with equal weight from analyses with two waveform models~\cite{GWTC-2.1,zenodo_GW190517}.
In the analysis of GW231123, and to a lesser extent GW190517, there is some uncertainty in the measurement of the source parameters due to systematics in the waveform modeling. Here, we mitigate this by using combined posteriors and indicating the constraints that are consistent across different waveform models (see Appendix~\ref{app:waveform_comp} for a detailed comparison).
The measured high spins of the constituent BHs in these two events make them particularly compelling for constraining ultralight boson masses through the superradiant instability.

Both events have strong support for being astrophysical in origin, consistent with binary BH mergers, with false alarm rates (minimum over all pipelines) of $<10^{-4}$/yr and $3\times10^{-4}$/yr for GW231123 and GW190517, respectively~\cite{GWTC-2.1,GW231123}. GW190517, in particular, has an astrophysical probability of almost unity. While GW231123 is a short-duration signal that could potentially arise from other burstlike astrophysical or cosmological sources, such as core-collapse supernovae, cosmic strings, or exotic compact objects, the binary BH merger interpretation remains the most probable astrophysical explanation~\cite{GW231123}. This study is carried out under the assumption that both events are quasicircular binary BH mergers. The network matched filter signal-to-noise ratios (SNRs) of GW231123 and GW190517 are $22.6^{+0.2}_{-0.3}$ and $10.8^{+0.5}_{-0.6}$, respectively~\cite{GWTC-2.1,GW231123}. The relatively high SNRs, especially for GW231123, compared to the detection threshold, yield well-constrained posterior distributions of the source properties that are sufficiently distinct from the priors. We also account for the potential impact of the prior distributions in our analysis. 
While large spins could, in principle, arise from noise fluctuations, the probability of obtaining two such events by chance among the $\lesssim 200$ detections to-date with well-measured source properties~\cite{LIGOScientific:2025slb} is modest and does not indicate a significant trials-factor concern. We also assume that the parameter estimation is not significantly affected by non-Gaussian noise (e.g., microglitches; see Refs.~\cite{Ray:2025rtt,Ashton:2021tvz}).

\ssec{Modeling black hole superradiance}%
If one considers a population of BHs that have spun down due to the presence of a scalar or vector boson with mass $\mb$ over a timescale longer than $T_{\rm age}$, there exists an excluded region in the BH mass-spin ($M$, $\chi$) plane, where all BHs would have been spun down via superradiance to below a maximum allowed spin, $\chi_{\rm max}(M,\mb, T_{\rm age})$.
We calculate $\chi_{\rm max}$ using the \texttt{SuperRad} package~\cite{Siemonsen:2022yyf,May2025}\footnote{\url{www.bitbucket.org/weast/superrad}}, assuming conservatively that the ultralight boson cloud grows from a single particle up to the mass where the superradiance saturates and the instability shuts off. The superradiant instability occurs when the oscillation frequency $\omega_R\approx \mb c^2/\hbar$ satisfies the condition $\omega_R<m \Omega_{\rm BH}$, where $m$ is the azimuthal number of the unstable mode, and $\Omega_{\rm BH}$ is the horizon frequency of the BH. 
This condition implies that higher-$m$ modes can be unstable and contribute to spin-down for higher boson masses or lower BH spins, including scenarios where the BH is spun down successively by multiple azimuthal modes. 
In this analysis, we use fully relativistic superradiant frequencies and instability rates for modes with $m \leq 5$ in the case of vector clouds, and $m\leq 2$ for scalars. For higher $m$ values beyond these ranges, we adopt nonrelativistic approximations. 
In particular, the superradiant instability is faster for the vector case, so higher $m$ modes are more relevant, and the nonrelativistic approximation can be less accurate when compared to scalars. See Appendices~\ref{app:spindown_calc} and~\ref{app:higher_m} for more details.

When considering bosons with self-interactions or interactions with other fields, these effects can cause the superradiant instability to saturate at a lower level, before the BH is fully spun down. Here, we consider several scalar and vector models and calculate the maximum strength of these interactions, in terms of the relevant coupling parameters, beyond which the spin-down constraints would be modified. In other words, we identify the parameter space where superradiance remains effective and the BH exclusion regions are unaffected by the extra interactions. We briefly summarize these below, with more details given in Appendix~\ref{app:interact}.

We consider an axion (scalar) model with an attractive quartic self-interaction $\mathcal{L} \supset -(m_b/f)^2 \phi^4/4!$. In this case, for sufficiently small energy scale $f$, the self-interaction term will cause bosonic radiation and leakage into modes which are absorbed by the BH. This effect can inhibit the growth of the superradiant cloud and suppress the efficiency of spin-down~\cite{Arvanitaki:2010sy,Baryakhtar:2020gao,Omiya:2022mwv,Witte:2024drg}. Here, we follow Ref.~\cite{Baryakhtar:2020gao} in bounding $f$. We neglect higher-order states, though including these could lead to improved constraints~\cite{Witte:2024drg}.

Additionally, we consider two types of dark photon (vector) interactions. The first is a small kinetic mixing between the massive dark photon and the Standard Model photon $\mathcal{L}\supset \varepsilon F'_{ab}F^{ab}/2$. For sufficiently large $\varepsilon$, the dark photon cloud can source a pair plasma which will dissipate energy through electromagnetic radiation, slowing down the growth of the cloud~\cite{Siemonsen:2022ivj} (see also Ref.~\cite{Xin:2024trp}). 
The second type of interaction is the Higgs-Abelian model.  A complex scalar with a Higgs-like potential $\mathcal{L} \supset |D^a\Phi|^2/2-\lambda/4(|\Phi|^2-v^2)^2$ is coupled to the dark photon through the gauge covariant derivative $D_a\equiv \nabla_a-ig A_a'$, giving the dark photon a 
mass $m_b=gv$, where $v$ is the vacuum expectation value of the scalar. In this case, for sufficiently large coupling strengths, the cloud can emit bosonic radiation~\cite{Fukuda:2019ewf,Jones:2024fpg}, or even form strings when it grows sufficiently large~\cite{East:2022ppo,East:2022rsi,Brzeminski:2024drp}, disrupting the exponential growth.
In setting bounds on these dark photon interactions, we follow Ref.~\cite{Jones:2024fpg}.

\ssec{Methods for setting constraints}
We now describe the procedure for setting constraints.
Given the uncertainties in the time elapsed between a BH's formation (or when it was last spun up, e.g., due to accretion) and its observation at merger, we consider a range of values for $T_{\rm age}$. As an upper bound, we adopt $10^7$ years, corresponding to a typical inspiral timescale in standard binary formation scenarios~\cite{PortegiesZwart:1999nm,Dominik:2013tma,Bavera:2020inc}. 
As a lower bound, we consider $10^5$ years, consistent with the merger timescales for dynamically formed binaries in dense environments such as globular clusters or active galactic nuclei~\cite{Rodriguez:2016kxx,Bartos:2016dgn,Yang:2019cbr}.
Unless otherwise specified, we use the lower bound when quoting conservative constraints.
We also note that we ignore any effects from the binary companion on the superradiant growth of a boson cloud, so $T_{\rm age}$ should be considered as the lower bound on the time prior to the binary reaching sufficiently small separations for such effects to become significant.

Given a boson of mass $\mb$ and an assumed age $T_{\rm age}$, for each set of inferred constituent BH masses and spins from compact binary coalescence parameter estimation, we evaluate whether the spins lie below the superradiance-imposed limit, $\chi_{\rm max}(M,\mb, T_{\rm age})$.
In principle, using the spins of both the BHs
in the binary will give the strongest constraints.
However, the spin of the secondary (less massive) constituent BH in GW190517 is not well-constrained~\cite{GWTC-2.1}, and the secondary spin in GW231123, though favored to be high, has a stronger dependence on the waveform model used~\cite{GW231123}. Therefore, here we only use the primary (more massive) constituent BHs in both of the merger events. 
More details about the constraints from adding the secondary BH of GW231123 with specific waveform models are given in Appendices~\ref{app:prior_vs_posterior} and \ref{app:waveform_comp}.

With the full set of multidimensional posterior samples for each merger event, we take every set of primary constituent BH mass and spin, denoted by $M_1^{(i)}$ and $\chi_1^{(i)}$, and compute the posterior-driven nonexclusion probability in a discrete form: 
\begin{equation}
   P(\mb, T_{\rm age}) = \frac{1}{N_{\mathrm{BH}}} \sum_{i=1}^{N_{\mathrm{BH}}} I{\left\{ \chi_1^{(i)} < \chi_{\max}(M_1^{(i)}, \mb, T_{\rm age}) \right\}},
   \label{eq:nonexclusion}
\end{equation}
where $N_{\rm BH}$ is the total number of posterior samples from the parameter estimation, and $I\left\{\cdot \right\}$ is an indicator function that returns unity if the primary BH spin lies below its respective superradiance-imposed upper bound $\chi_{\max}$, and zero otherwise. 
This expression defines the probability $P(\mb, T_{\rm age})$ that a given boson mass $\mb$ and an assumed age $T_{\rm age}$ are consistent with the observed primary constituent BH.

One potential issue is that, in some parts of the parameter space, $\chi_{\rm max}$ can be small enough to be inconsistent with a sizable fraction of even the prior probability
distribution, so it is important to check that any exclusion is being driven by the data. To mitigate the potential impacts introduced by the uniform prior spin distribution over the range of $[0,1)$, we compute the spin prior-driven nonexclusion probability, $P'(\mb, T_{\rm age})$, following Eq.~\eqref{eq:nonexclusion} but drawing $\chi_1$ samples from a uniform distribution. 
We then exclude the boson mass $\mb$ at a 90\% confidence level for a given spin-down timescale $T_{\rm age}$ if 
$P(\mb, T_{\rm age}) < 0.1 P'(\mb, T_{\rm age})$.  
This criterion is more restrictive than $P(\mb, T_{\rm age}) < 0.1$ and
requires that the posterior-driven nonexclusion probability decreases by more than a factor of 10 compared to the prior-driven result, ensuring that the constraint arises from the data rather than just the uniform spin prior.
See Appendix~\ref{app:prior_vs_posterior} for more details.

Typically, the full posterior distributions obtained from parameter estimation contain $N_{BH} \sim \mathcal{O}(10^4)$ samples, which sets a limit on the statistical precision of the exclusion confidence to $\gtrsim 0.01$.
We repeat the calculation on a grid of $(\mb, T_{\rm age})$ values, for a mass range spanning $\sim 10^{-15}$--$10^{-11}$~{\rm eV} and a spin-down timescale of $10^5$--$10^7$ years.

\begin{figure}[htbp]
    \centering
    \includegraphics[width=\linewidth]{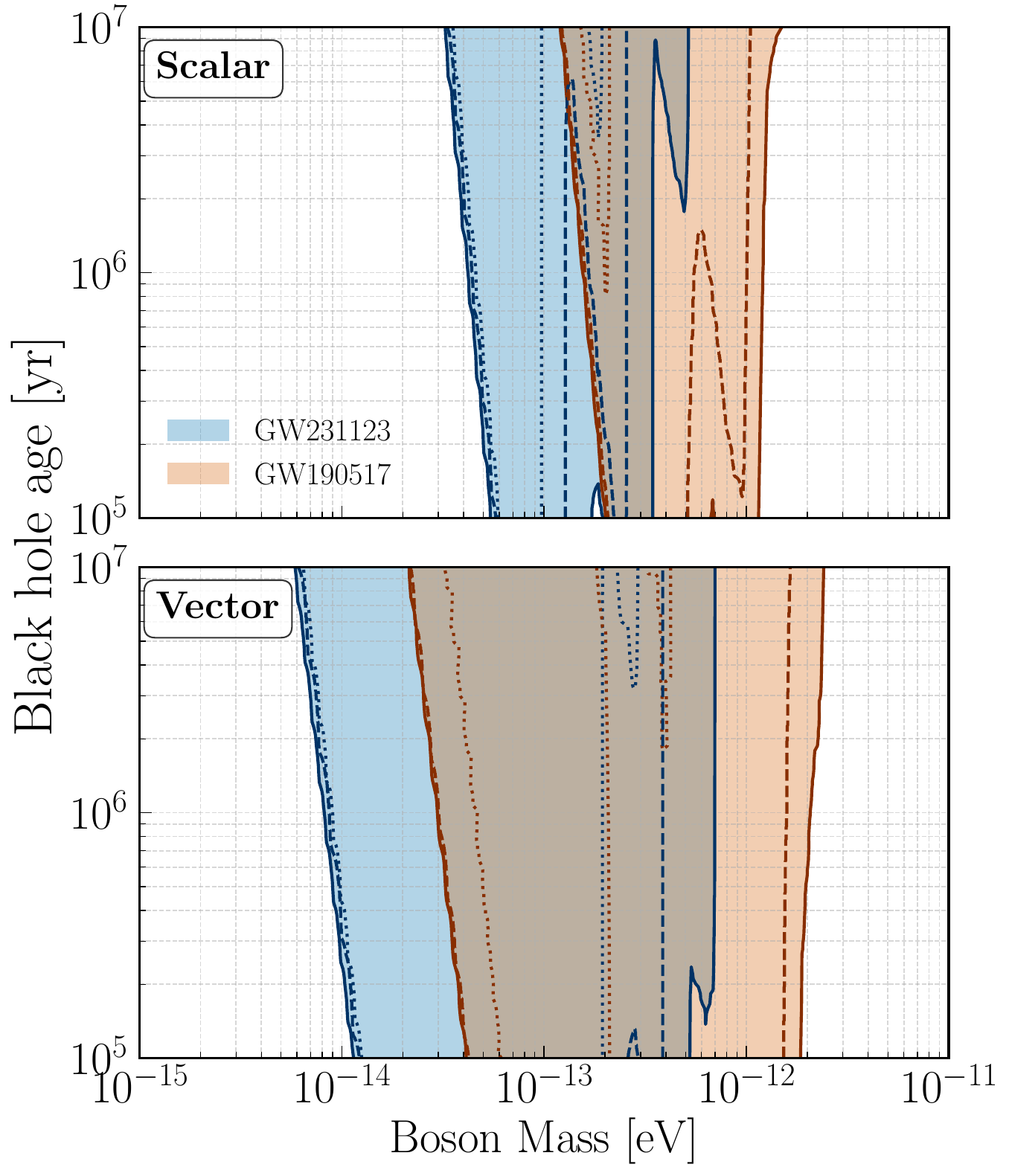}
    \caption{Exclusion regions for scalar and vector boson masses as a function of the BH age. 
    The blue and orange shaded regions are excluded by GW231123 and GW190517, respectively, at confidence levels above 90\% (i.e., $P<0.1P'$).
    The regions enclosed by dotted curves correspond to confidence levels above 99\% (i.e., $P<0.01P'$), while the dashed curves mark the conservative exclusion regions common to all waveform models at a confidence level above 90\%.
    }
    \label{fig:confidence_plot}
\end{figure}

\ssec{Constraints}%
Our results for GW231123 and GW190517 are presented in Fig.~\ref{fig:confidence_plot}. 
The exclusion regions indicate boson masses that are incompatible with the observed BH spins at 90\% confidence, assuming superradiant spin-down over the specified timescale.
We perform separate analyses for scalar and vector bosons for each event. 
Assuming a conservative BH age of $10^{5}$ years, we exclude scalar bosons in the mass range of \heavyBBHscalarlimits~eV for GW231123 and \lightBBHscalarlimits~eV for GW190517 at 90\% confidence. 
For vector bosons, the excluded mass ranges are broader, \heavyBBHvectorlimits~eV for GW231123 and \lightBBHvectorlimits~eV for GW190517, reflecting the faster superradiant growth rates for vector fields. 
The exclusion region expands with increasing $T_{\rm age}$, since older BHs would have experienced a longer duration for superradiant spin-down, thereby strengthening the constraints on boson masses, though the dependence is not strong in this range. As can also been seen from Fig.~\ref{fig:confidence_plot}, the highest confidence exclusions (indicated by the 99\% confidence contours) are in the low boson mass region where the maximum allowed spin is smallest. 
This lower mass region is also the most robust to waveform-dependent systematics, as indicated by the dashed curves in Fig.~\ref{fig:confidence_plot} enclosing the excluded regions common to all waveform models when considering their posteriors individually;
see also Appendix~\ref{app:waveform_comp}.

The disfavored mass ranges obtained from GW190517 are at noticeably higher boson masses as a result of the lower mass of the primary constituent BH and the corresponding shift in the superradiance condition.
These results are obtained by combining posterior samples from multiple waveform models. However, the constraints derived from the posterior samples of each individual waveform model are largely consistent with one another.
These results are also obtained using the primary BH alone for both events. If we include the secondary BH for GW231123 with waveform models that infer a well-constrained high-spin secondary, the upper bound on the exclusion increases to approximately $8 \times 10^{-13}$ eV and $1 \times 10^{-12}$ eV for the scalar and vector bosons, respectively. See Appendix~\ref{app:waveform_comp} for details.

\begin{figure*}[htbp]
    \centering
\includegraphics[width=\linewidth]{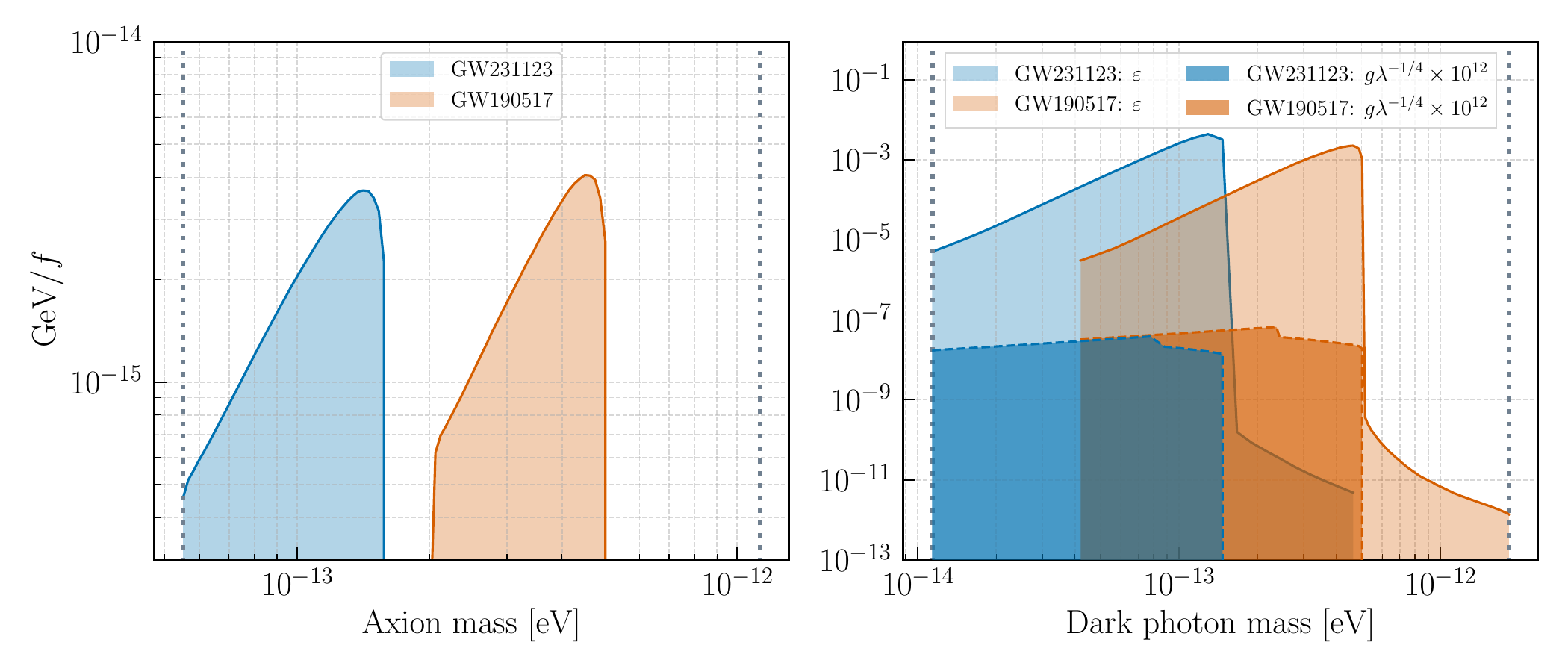}

    \caption{Excluded parameter space for coupled bosons from GW231123 and GW190517. 
    The shaded regions (down to vanishing coupling strength) show the parameter space excluded at the 90\% confidence level, assuming a spin-down timescale of $T_{\rm age}=10^5$ years; i.e., we identified the regions, where the couplings are sufficiently weak so as to not affect BH spin-down via superradiance. 
    \textit{Left}: Constraints on the inverse energy scale $f^{-1}$ for an axion with an attractive quartic self-interaction.
    \textit{Right}: Constraints on the dark photon parameter space with a kinetic mixing parameter $\varepsilon$ (light blue and orange regions) and in the Higgs-Abelian model (dark blue and orange regions). 
    For the latter, the constraints on the combination of dimensionless coupling parameters $g\lambda^{-1/4}$ have been scaled up by a factor of $10^{12}$ to be accommodated on the plot. Due to theoretical uncertainties, we do not show bounds from superradiance involving higher azimuthal numbers ($m>1$), except for the kinetically mixed dark photon, in which case we use much less constraining assumptions for $m>1$. 
    With improved theoretical modeling of the interaction effects, it should be possible to extend the constraints to cover the entire boson mass region excluded under the assumption of purely gravitational interactions (as indicated by the dotted vertical lines).
    }
    \label{fig:coupled_bosons}
\end{figure*}

The above constraints assume purely gravitational interactions. In Fig.~\ref{fig:coupled_bosons}, we extend the analysis to boson models with additional interactions, restricting to a fixed spin-down timescale of $T_{\rm age}=10^5$ years. 
We find that these spin-down constraints apply to axion models with energy scale $f \gtrsim10^{15}$ GeV, which includes the decay constant of the QCD axion within the relevant mass range~\cite{Peccei:1977hh,Weinberg:1977ma,Wilczek:1977pj}.
For a kinetically mixed dark photon, the spin-down bounds exclude values of the mixing parameter $\varepsilon$ that, in most of the constrained parameter space, extend beyond the maximum allowed value $\varepsilon \lesssim 10^{-6}$ set by cosmic microwave background observations~\cite{Mirizzi:2009iz,Caputo:2020bdy,Fixsen:1996nj,McCarthy:2024ozh,Pirvu:2023lch,Aramburo-Garcia:2024cbz} (see also Ref.~\cite{Mirasola:2025car}). Together, these constraints effectively rule out any value of $\varepsilon$ in this mass range.
In the Higgs-Abelian model, the couplings must be very small for superradiant spin-down to remain unaffected, with $g\lambda^{-1/4} \lesssim 10^{-20}$. In terms of an approximate kinetic mixing parameter generated from heavy fermion loops with order-unity charge \cite{Holdom1986}, this corresponds to $\varepsilon \lesssim 10^{-23} \lambda^{1/4}$.

Considering larger values for the BH age decreases the lower bound on the axion energy scale where the exclusions apply (while slightly broadening the mass range) with $f\propto T_{\rm age}^{-1/2}$. The dark photon constraints remain largely unchanged over the age range of $10^5$ to $10^7$ years since, in this analysis (unlike in the axion case), we do not consider scenarios where the boson cloud extracts angular momentum from the BH at a slower-than-exponential growth rate. 
In deriving these constraints, we have been conservative due to the lack of a complete theoretical understanding of interaction effects, in particular for higher azimuthal number modes, which are relevant for larger boson masses.

\ssec{Discussion and conclusion}
With two complementary GW observations, GW231123 and GW190517, we exclude scalar bosons with masses in the range of \scalarlimits~eV and vector bosons in the range of \vectorlimits~eV at 90\% confidence, assuming a conservative BH lifetime over $10^5$ years. 
These excluded boson mass ranges extend to lower values than previous observations, driven in large part by the fact that the primary BH in GW231123 has an unusually large mass, above $100\ M_{\odot}$. 
 We also note that the lower boson mass range is the highest confidence exclusion and the most robust, e.g., to different waveform choices.
In comparison, previous analyses using electromagnetic observations of accreting stellar mass BHs have lower bounds for their excluded regions of $\sim3\times10^{-13}$ eV
for scalars~\cite{Baryakhtar:2020gao,Hoof:2024quk}, and $\sim 5\times10^{-14}$ eV for vectors~\cite{Baryakhtar:2017ngi}.

The exclusion region we find for scalars using GW190517 is similar to that found in Ref.~\cite{Ng:2020ruv} when using the same age assumptions (falling
somewhere between 90\% and 99\% regions in Fig.~\ref{fig:confidence_plot}).
That reference measures the overall spin distribution of the binary BH population and the scalar boson mass simultaneously, which has the advantage that it can also be used to find evidence for the existence of scalars.
Here, we focus exclusively on deriving constraints using the parameter estimation with standard priors of specific events. The lower bound of the excluded mass range for vectors found here using GW190517 is also comparable to that in Ref.~\cite{Ghosh:2021zuf}, though here we exclude a larger range by including higher azimuthal modes.

One key assumption throughout this analysis is that GW231123 and GW190517 are quasicircular binary BH mergers. While both events have high astrophysical probabilities consistent with the standard binary BH merger interpretation, alternative burstlike astrophysical scenarios cannot be completely ruled out. 

We have also derived constraints on models with bosonic self-interactions or couplings to other fields, when these interactions are sufficiently weak so as not to disrupt superradiant spin-down. This includes axions (like the QCD axion) with energy scale $f\gtrsim10^{15}$~GeV and dark photons with kinetic mixing $\varepsilon<10^{-6}$. The constraints derived here are conservative due to theoretical uncertainties. They can likely be extended in the higher mass range, excluding nongravitational interactions
purely through theoretical progress, in particular by better modeling the interaction effects for higher azimuthal number modes.

\noindent\textbf{Note added:} Shortly after this work, Ref.~\cite{Caputo:2025oap} appeared, analyzing GW231123 and placing constraints on axionlike particles subject to self-interaction with scale $f$. 
Broadly our results are consistent. Particularly, the constraints on the scalar mass are comparable, while their constraints on $f$ are somewhat stronger for this event. This difference may be due to the fact that
Ref.~\cite{Caputo:2025oap} includes higher-order modes and explicit time evolution for axion self-interactions (while we follow Ref.~\cite{Baryakhtar:2020gao} and restrict to $GM m_b/(c\hbar)<0.2$). That study also uses both the primary and secondary BH spins from a single waveform posterior.

\acknowledgments
The authors would like to thank Leo Tsukada for the helpful discussion and comments during the internal review.
This material is based upon work supported by NSF's LIGO Laboratory which is a major facility fully funded by the National Science Foundation.
The authors are grateful for computational resources provided by the LIGO Laboratory and supported by National Science Foundation Grants PHY--0757058 and PHY--0823459.
This research is supported by the Australian Research Council Centre of Excellence for Gravitational Wave Discovery (OzGrav), Project Number CE230100016. 
L.S. is also supported by the Australian Research Council Discovery Early Career Researcher Award, Project Number DE240100206. 
W.E. acknowledges support from a Natural Sciences and Engineering Research
Council of Canada Discovery Grant and
an Ontario Ministry of Colleges and Universities Early Researcher Award.
This research was 
supported in part by Perimeter Institute for Theoretical Physics. Research at
Perimeter Institute is supported in part by the Government of Canada through
the Department of Innovation, Science and Economic Development and by the
Province of Ontario through the Ministry of Colleges and Universities.
This manuscript carries LIGO Document No. DCC--P2500452.\\

\def\bibsection{\section*{References}}
\bibliography{refs}

\begin{thebibliography}{92}%
\makeatletter
\providecommand \@ifxundefined [1]{%
 \@ifx{#1\undefined}
}%
\providecommand \@ifnum [1]{%
 \ifnum #1\expandafter \@firstoftwo
 \else \expandafter \@secondoftwo
 \fi
}%
\providecommand \@ifx [1]{%
 \ifx #1\expandafter \@firstoftwo
 \else \expandafter \@secondoftwo
 \fi
}%
\providecommand \natexlab [1]{#1}%
\providecommand \enquote  [1]{``#1''}%
\providecommand \bibnamefont  [1]{#1}%
\providecommand \bibfnamefont [1]{#1}%
\providecommand \citenamefont [1]{#1}%
\providecommand \href@noop [0]{\@secondoftwo}%
\providecommand \href [0]{\begingroup \@sanitize@url \@href}%
\providecommand \@href[1]{\@@startlink{#1}\@@href}%
\providecommand \@@href[1]{\endgroup#1\@@endlink}%
\providecommand \@sanitize@url [0]{\catcode `\\12\catcode `\$12\catcode
  `\&12\catcode `\#12\catcode `\^12\catcode `\_12\catcode `\%12\relax}%
\providecommand \@@startlink[1]{}%
\providecommand \@@endlink[0]{}%
\providecommand \url  [0]{\begingroup\@sanitize@url \@url }%
\providecommand \@url [1]{\endgroup\@href {#1}{\urlprefix }}%
\providecommand \urlprefix  [0]{URL }%
\providecommand \Eprint [0]{\href }%
\providecommand \doibase [0]{http://dx.doi.org/}%
\providecommand \selectlanguage [0]{\@gobble}%
\providecommand \bibinfo  [0]{\@secondoftwo}%
\providecommand \bibfield  [0]{\@secondoftwo}%
\providecommand \translation [1]{[#1]}%
\providecommand \BibitemOpen [0]{}%
\providecommand \bibitemStop [0]{}%
\providecommand \bibitemNoStop [0]{.\EOS\space}%
\providecommand \EOS [0]{\spacefactor3000\relax}%
\providecommand \BibitemShut  [1]{\csname bibitem#1\endcsname}%
\let\auto@bib@innerbib\@empty
\bibitem [{\citenamefont {Peccei}\ and\ \citenamefont
  {Quinn}(1977{\natexlab{a}})}]{Peccei:1977hh}%
  \BibitemOpen
  \bibfield  {author} {\bibinfo {author} {\bibfnamefont {R.~D.}\ \bibnamefont
  {Peccei}}\ and\ \bibinfo {author} {\bibfnamefont {H.~R.}\ \bibnamefont
  {Quinn}},\ }\href {\doibase 10.1103/PhysRevLett.38.1440} {\bibfield
  {journal} {\bibinfo  {journal} {Phys. Rev. Lett.}\ }\textbf {\bibinfo
  {volume} {38}},\ \bibinfo {pages} {1440} (\bibinfo {year}
  {1977}{\natexlab{a}})}\BibitemShut {NoStop}%
\bibitem [{\citenamefont {Weinberg}(1978)}]{Weinberg:1977ma}%
  \BibitemOpen
  \bibfield  {author} {\bibinfo {author} {\bibfnamefont {S.}~\bibnamefont
  {Weinberg}},\ }\href {\doibase 10.1103/PhysRevLett.40.223} {\bibfield
  {journal} {\bibinfo  {journal} {Phys. Rev. Lett.}\ }\textbf {\bibinfo
  {volume} {40}},\ \bibinfo {pages} {223} (\bibinfo {year} {1978})}\BibitemShut
  {NoStop}%
\bibitem [{\citenamefont {Peccei}\ and\ \citenamefont
  {Quinn}(1977{\natexlab{b}})}]{Peccei:1977ur}%
  \BibitemOpen
  \bibfield  {author} {\bibinfo {author} {\bibfnamefont {R.~D.}\ \bibnamefont
  {Peccei}}\ and\ \bibinfo {author} {\bibfnamefont {H.~R.}\ \bibnamefont
  {Quinn}},\ }\href {\doibase 10.1103/PhysRevD.16.1791} {\bibfield  {journal}
  {\bibinfo  {journal} {Phys. Rev. D}\ }\textbf {\bibinfo {volume} {16}},\
  \bibinfo {pages} {1791} (\bibinfo {year} {1977}{\natexlab{b}})}\BibitemShut
  {NoStop}%
\bibitem [{\citenamefont {Arvanitaki}\ \emph {et~al.}(2010)\citenamefont
  {Arvanitaki}, \citenamefont {Dimopoulos}, \citenamefont {Dubovsky},
  \citenamefont {Kaloper},\ and\ \citenamefont
  {March-Russell}}]{Arvanitaki:2009fg}%
  \BibitemOpen
  \bibfield  {author} {\bibinfo {author} {\bibfnamefont {A.}~\bibnamefont
  {Arvanitaki}}, \bibinfo {author} {\bibfnamefont {S.}~\bibnamefont
  {Dimopoulos}}, \bibinfo {author} {\bibfnamefont {S.}~\bibnamefont
  {Dubovsky}}, \bibinfo {author} {\bibfnamefont {N.}~\bibnamefont {Kaloper}}, \
  and\ \bibinfo {author} {\bibfnamefont {J.}~\bibnamefont {March-Russell}},\
  }\href {\doibase 10.1103/PhysRevD.81.123530} {\bibfield  {journal} {\bibinfo
  {journal} {Phys. Rev. D}\ }\textbf {\bibinfo {volume} {81}},\ \bibinfo
  {pages} {123530} (\bibinfo {year} {2010})},\ \Eprint
  {http://arxiv.org/abs/0905.4720} {arXiv:0905.4720 [hep-th]} \BibitemShut
  {NoStop}%
\bibitem [{\citenamefont {Goodsell}\ \emph {et~al.}(2009)\citenamefont
  {Goodsell}, \citenamefont {Jaeckel}, \citenamefont {Redondo},\ and\
  \citenamefont {Ringwald}}]{Goodsell:2009xc}%
  \BibitemOpen
  \bibfield  {author} {\bibinfo {author} {\bibfnamefont {M.}~\bibnamefont
  {Goodsell}}, \bibinfo {author} {\bibfnamefont {J.}~\bibnamefont {Jaeckel}},
  \bibinfo {author} {\bibfnamefont {J.}~\bibnamefont {Redondo}}, \ and\
  \bibinfo {author} {\bibfnamefont {A.}~\bibnamefont {Ringwald}},\ }\href
  {\doibase 10.1088/1126-6708/2009/11/027} {\bibfield  {journal} {\bibinfo
  {journal} {JHEP}\ }\textbf {\bibinfo {volume} {11}},\ \bibinfo {pages} {027}
  (\bibinfo {year} {2009})},\ \Eprint {http://arxiv.org/abs/0909.0515}
  {arXiv:0909.0515 [hep-ph]} \BibitemShut {NoStop}%
\bibitem [{\citenamefont {Jaeckel}\ and\ \citenamefont
  {Ringwald}(2010)}]{Jaeckel:2010ni}%
  \BibitemOpen
  \bibfield  {author} {\bibinfo {author} {\bibfnamefont {J.}~\bibnamefont
  {Jaeckel}}\ and\ \bibinfo {author} {\bibfnamefont {A.}~\bibnamefont
  {Ringwald}},\ }\href {\doibase 10.1146/annurev.nucl.012809.104433} {\bibfield
   {journal} {\bibinfo  {journal} {Ann. Rev. Nucl. Part. Sci.}\ }\textbf
  {\bibinfo {volume} {60}},\ \bibinfo {pages} {405} (\bibinfo {year} {2010})},\
  \Eprint {http://arxiv.org/abs/1002.0329} {arXiv:1002.0329 [hep-ph]}
  \BibitemShut {NoStop}%
\bibitem [{\citenamefont {Essig}\ \emph {et~al.}(2013)\citenamefont {Essig}
  \emph {et~al.}}]{Essig:2013lka}%
  \BibitemOpen
  \bibfield  {author} {\bibinfo {author} {\bibfnamefont {R.}~\bibnamefont
  {Essig}} \emph {et~al.},\ }in\ \href@noop {} {\emph {\bibinfo {booktitle}
  {{Snowmass 2013}: {Snowmass on the Mississippi}}}}\ (\bibinfo {year} {2013})\
  \Eprint {http://arxiv.org/abs/1311.0029} {arXiv:1311.0029 [hep-ph]}
  \BibitemShut {NoStop}%
\bibitem [{\citenamefont {Hui}\ \emph {et~al.}(2017)\citenamefont {Hui},
  \citenamefont {Ostriker}, \citenamefont {Tremaine},\ and\ \citenamefont
  {Witten}}]{Hui:2016ltb}%
  \BibitemOpen
  \bibfield  {author} {\bibinfo {author} {\bibfnamefont {L.}~\bibnamefont
  {Hui}}, \bibinfo {author} {\bibfnamefont {J.~P.}\ \bibnamefont {Ostriker}},
  \bibinfo {author} {\bibfnamefont {S.}~\bibnamefont {Tremaine}}, \ and\
  \bibinfo {author} {\bibfnamefont {E.}~\bibnamefont {Witten}},\ }\href
  {\doibase 10.1103/PhysRevD.95.043541} {\bibfield  {journal} {\bibinfo
  {journal} {Phys. Rev. D}\ }\textbf {\bibinfo {volume} {95}},\ \bibinfo
  {pages} {043541} (\bibinfo {year} {2017})},\ \Eprint
  {http://arxiv.org/abs/1610.08297} {arXiv:1610.08297 [astro-ph.CO]}
  \BibitemShut {NoStop}%
\bibitem [{\citenamefont {Brito}\ \emph
  {et~al.}(2015{\natexlab{a}})\citenamefont {Brito}, \citenamefont {Cardoso},\
  and\ \citenamefont {Pani}}]{Brito:2015rjv}%
  \BibitemOpen
  \bibfield  {author} {\bibinfo {author} {\bibfnamefont {R.}~\bibnamefont
  {Brito}}, \bibinfo {author} {\bibfnamefont {V.}~\bibnamefont {Cardoso}}, \
  and\ \bibinfo {author} {\bibfnamefont {P.}~\bibnamefont {Pani}},\ }\href
  {\doibase 10.1007/978-3-319-19000-6_3} {\bibfield  {journal} {\bibinfo
  {journal} {Lect. Notes Phys.}\ }\textbf {\bibinfo {volume} {906}},\ \bibinfo
  {pages} {35} (\bibinfo {year} {2015}{\natexlab{a}})}\BibitemShut {NoStop}%
\bibitem [{\citenamefont {Starobinsky}(1973)}]{Starobinsky:1973aij}%
  \BibitemOpen
  \bibfield  {author} {\bibinfo {author} {\bibfnamefont {A.~A.}\ \bibnamefont
  {Starobinsky}},\ }\href@noop {} {\bibfield  {journal} {\bibinfo  {journal}
  {Sov. Phys. JETP}\ }\textbf {\bibinfo {volume} {37}},\ \bibinfo {pages} {28}
  (\bibinfo {year} {1973})}\BibitemShut {NoStop}%
\bibitem [{\citenamefont {{Zel'Dovich}}(1971)}]{1971JETPL..14..180Z}%
  \BibitemOpen
  \bibfield  {author} {\bibinfo {author} {\bibfnamefont {Y.~B.}\ \bibnamefont
  {{Zel'Dovich}}},\ }\href@noop {} {\bibfield  {journal} {\bibinfo  {journal}
  {Soviet Journal of Experimental and Theoretical Physics Letters}\ }\textbf
  {\bibinfo {volume} {14}},\ \bibinfo {pages} {180} (\bibinfo {year}
  {1971})}\BibitemShut {NoStop}%
\bibitem [{\citenamefont {Misner}(1972)}]{PhysRevLett.28.994}%
  \BibitemOpen
  \bibfield  {author} {\bibinfo {author} {\bibfnamefont {C.~W.}\ \bibnamefont
  {Misner}},\ }\href {\doibase 10.1103/PhysRevLett.28.994} {\bibfield
  {journal} {\bibinfo  {journal} {Phys. Rev. Lett.}\ }\textbf {\bibinfo
  {volume} {28}},\ \bibinfo {pages} {994} (\bibinfo {year} {1972})}\BibitemShut
  {NoStop}%
\bibitem [{\citenamefont {Detweiler}(1980)}]{Detweiler:1980uk}%
  \BibitemOpen
  \bibfield  {author} {\bibinfo {author} {\bibfnamefont {S.~L.}\ \bibnamefont
  {Detweiler}},\ }\href {\doibase 10.1103/PhysRevD.22.2323} {\bibfield
  {journal} {\bibinfo  {journal} {Phys. Rev. D}\ }\textbf {\bibinfo {volume}
  {22}},\ \bibinfo {pages} {2323} (\bibinfo {year} {1980})}\BibitemShut
  {NoStop}%
\bibitem [{\citenamefont {Press}\ and\ \citenamefont
  {Teukolsky}(1972)}]{Press:1972zz}%
  \BibitemOpen
  \bibfield  {author} {\bibinfo {author} {\bibfnamefont {W.~H.}\ \bibnamefont
  {Press}}\ and\ \bibinfo {author} {\bibfnamefont {S.~A.}\ \bibnamefont
  {Teukolsky}},\ }\href {\doibase 10.1038/238211a0} {\bibfield  {journal}
  {\bibinfo  {journal} {Nature}\ }\textbf {\bibinfo {volume} {238}},\ \bibinfo
  {pages} {211} (\bibinfo {year} {1972})}\BibitemShut {NoStop}%
\bibitem [{\citenamefont {Bekenstein}(1973)}]{Bekenstein:1973mi}%
  \BibitemOpen
  \bibfield  {author} {\bibinfo {author} {\bibfnamefont {J.~D.}\ \bibnamefont
  {Bekenstein}},\ }\href {\doibase 10.1103/PhysRevD.7.949} {\bibfield
  {journal} {\bibinfo  {journal} {Phys. Rev. D}\ }\textbf {\bibinfo {volume}
  {7}},\ \bibinfo {pages} {949} (\bibinfo {year} {1973})}\BibitemShut {NoStop}%
\bibitem [{\citenamefont {Arvanitaki}\ and\ \citenamefont
  {Dubovsky}(2011)}]{Arvanitaki:2010sy}%
  \BibitemOpen
  \bibfield  {author} {\bibinfo {author} {\bibfnamefont {A.}~\bibnamefont
  {Arvanitaki}}\ and\ \bibinfo {author} {\bibfnamefont {S.}~\bibnamefont
  {Dubovsky}},\ }\href {\doibase 10.1103/PhysRevD.83.044026} {\bibfield
  {journal} {\bibinfo  {journal} {Phys. Rev. D}\ }\textbf {\bibinfo {volume}
  {83}},\ \bibinfo {pages} {044026} (\bibinfo {year} {2011})},\ \Eprint
  {http://arxiv.org/abs/1004.3558} {arXiv:1004.3558 [hep-th]} \BibitemShut
  {NoStop}%
\bibitem [{\citenamefont {Furuhashi}\ and\ \citenamefont
  {Nambu}(2004)}]{Furuhashi:2004jk}%
  \BibitemOpen
  \bibfield  {author} {\bibinfo {author} {\bibfnamefont {H.}~\bibnamefont
  {Furuhashi}}\ and\ \bibinfo {author} {\bibfnamefont {Y.}~\bibnamefont
  {Nambu}},\ }\href {\doibase 10.1143/PTP.112.983} {\bibfield  {journal}
  {\bibinfo  {journal} {Prog. Theor. Phys.}\ }\textbf {\bibinfo {volume}
  {112}},\ \bibinfo {pages} {983} (\bibinfo {year} {2004})},\ \Eprint
  {http://arxiv.org/abs/gr-qc/0402037} {arXiv:gr-qc/0402037} \BibitemShut
  {NoStop}%
\bibitem [{\citenamefont {Dolan}(2007)}]{Dolan:2007mj}%
  \BibitemOpen
  \bibfield  {author} {\bibinfo {author} {\bibfnamefont {S.~R.}\ \bibnamefont
  {Dolan}},\ }\href {\doibase 10.1103/PhysRevD.76.084001} {\bibfield  {journal}
  {\bibinfo  {journal} {Phys. Rev. D}\ }\textbf {\bibinfo {volume} {76}},\
  \bibinfo {pages} {084001} (\bibinfo {year} {2007})},\ \Eprint
  {http://arxiv.org/abs/0705.2880} {arXiv:0705.2880 [gr-qc]} \BibitemShut
  {NoStop}%
\bibitem [{\citenamefont {Brito}\ \emph
  {et~al.}(2015{\natexlab{b}})\citenamefont {Brito}, \citenamefont {Cardoso},\
  and\ \citenamefont {Pani}}]{Brito:2014wla}%
  \BibitemOpen
  \bibfield  {author} {\bibinfo {author} {\bibfnamefont {R.}~\bibnamefont
  {Brito}}, \bibinfo {author} {\bibfnamefont {V.}~\bibnamefont {Cardoso}}, \
  and\ \bibinfo {author} {\bibfnamefont {P.}~\bibnamefont {Pani}},\ }\href
  {\doibase 10.1088/0264-9381/32/13/134001} {\bibfield  {journal} {\bibinfo
  {journal} {Class. Quant. Grav.}\ }\textbf {\bibinfo {volume} {32}},\ \bibinfo
  {pages} {134001} (\bibinfo {year} {2015}{\natexlab{b}})},\ \Eprint
  {http://arxiv.org/abs/1411.0686} {arXiv:1411.0686 [gr-qc]} \BibitemShut
  {NoStop}%
\bibitem [{\citenamefont {East}\ and\ \citenamefont
  {Pretorius}(2017)}]{East:2017ovw}%
  \BibitemOpen
  \bibfield  {author} {\bibinfo {author} {\bibfnamefont {W.~E.}\ \bibnamefont
  {East}}\ and\ \bibinfo {author} {\bibfnamefont {F.}~\bibnamefont
  {Pretorius}},\ }\href {\doibase 10.1103/PhysRevLett.119.041101} {\bibfield
  {journal} {\bibinfo  {journal} {Phys. Rev. Lett.}\ }\textbf {\bibinfo
  {volume} {119}},\ \bibinfo {pages} {041101} (\bibinfo {year} {2017})},\
  \Eprint {http://arxiv.org/abs/1704.04791} {arXiv:1704.04791 [gr-qc]}
  \BibitemShut {NoStop}%
\bibitem [{\citenamefont {East}(2018)}]{East:2018glu}%
  \BibitemOpen
  \bibfield  {author} {\bibinfo {author} {\bibfnamefont {W.~E.}\ \bibnamefont
  {East}},\ }\href {\doibase 10.1103/PhysRevLett.121.131104} {\bibfield
  {journal} {\bibinfo  {journal} {Phys. Rev. Lett.}\ }\textbf {\bibinfo
  {volume} {121}},\ \bibinfo {pages} {131104} (\bibinfo {year} {2018})},\
  \Eprint {http://arxiv.org/abs/1807.00043} {arXiv:1807.00043 [gr-qc]}
  \BibitemShut {NoStop}%
\bibitem [{\citenamefont {Arvanitaki}\ \emph {et~al.}(2015)\citenamefont
  {Arvanitaki}, \citenamefont {Baryakhtar},\ and\ \citenamefont
  {Huang}}]{Arvanitaki:2014wva}%
  \BibitemOpen
  \bibfield  {author} {\bibinfo {author} {\bibfnamefont {A.}~\bibnamefont
  {Arvanitaki}}, \bibinfo {author} {\bibfnamefont {M.}~\bibnamefont
  {Baryakhtar}}, \ and\ \bibinfo {author} {\bibfnamefont {X.}~\bibnamefont
  {Huang}},\ }\href {\doibase 10.1103/PhysRevD.91.084011} {\bibfield  {journal}
  {\bibinfo  {journal} {Phys. Rev. D}\ }\textbf {\bibinfo {volume} {91}},\
  \bibinfo {pages} {084011} (\bibinfo {year} {2015})},\ \Eprint
  {http://arxiv.org/abs/1411.2263} {arXiv:1411.2263 [hep-ph]} \BibitemShut
  {NoStop}%
\bibitem [{\citenamefont {Stott}\ and\ \citenamefont
  {Marsh}(2018)}]{Stott:2018opm}%
  \BibitemOpen
  \bibfield  {author} {\bibinfo {author} {\bibfnamefont {M.~J.}\ \bibnamefont
  {Stott}}\ and\ \bibinfo {author} {\bibfnamefont {D.~J.~E.}\ \bibnamefont
  {Marsh}},\ }\href {\doibase 10.1103/PhysRevD.98.083006} {\bibfield  {journal}
  {\bibinfo  {journal} {Phys. Rev. D}\ }\textbf {\bibinfo {volume} {98}},\
  \bibinfo {pages} {083006} (\bibinfo {year} {2018})},\ \Eprint
  {http://arxiv.org/abs/1805.02016} {arXiv:1805.02016 [hep-ph]} \BibitemShut
  {NoStop}%
\bibitem [{\citenamefont {Baryakhtar}\ \emph {et~al.}(2021)\citenamefont
  {Baryakhtar}, \citenamefont {Galanis}, \citenamefont {Lasenby},\ and\
  \citenamefont {Simon}}]{Baryakhtar:2020gao}%
  \BibitemOpen
  \bibfield  {author} {\bibinfo {author} {\bibfnamefont {M.}~\bibnamefont
  {Baryakhtar}}, \bibinfo {author} {\bibfnamefont {M.}~\bibnamefont {Galanis}},
  \bibinfo {author} {\bibfnamefont {R.}~\bibnamefont {Lasenby}}, \ and\
  \bibinfo {author} {\bibfnamefont {O.}~\bibnamefont {Simon}},\ }\href
  {\doibase 10.1103/PhysRevD.103.095019} {\bibfield  {journal} {\bibinfo
  {journal} {Phys. Rev. D}\ }\textbf {\bibinfo {volume} {103}},\ \bibinfo
  {pages} {095019} (\bibinfo {year} {2021})},\ \Eprint
  {http://arxiv.org/abs/2011.11646} {arXiv:2011.11646 [hep-ph]} \BibitemShut
  {NoStop}%
\bibitem [{\citenamefont {Hoof}\ \emph {et~al.}(2024)\citenamefont {Hoof},
  \citenamefont {Marsh}, \citenamefont {Sisk-Reyn{\'e}s}, \citenamefont
  {Matthews},\ and\ \citenamefont {Reynolds}}]{Hoof:2024quk}%
  \BibitemOpen
  \bibfield  {author} {\bibinfo {author} {\bibfnamefont {S.}~\bibnamefont
  {Hoof}}, \bibinfo {author} {\bibfnamefont {D.~J.~E.}\ \bibnamefont {Marsh}},
  \bibinfo {author} {\bibfnamefont {J.}~\bibnamefont {Sisk-Reyn{\'e}s}},
  \bibinfo {author} {\bibfnamefont {J.~H.}\ \bibnamefont {Matthews}}, \ and\
  \bibinfo {author} {\bibfnamefont {C.}~\bibnamefont {Reynolds}},\ }\href@noop
  {} {\  (\bibinfo {year} {2024})},\ \Eprint {http://arxiv.org/abs/2406.10337}
  {arXiv:2406.10337 [hep-ph]} \BibitemShut {NoStop}%
\bibitem [{\citenamefont {Baryakhtar}\ \emph {et~al.}(2017)\citenamefont
  {Baryakhtar}, \citenamefont {Lasenby},\ and\ \citenamefont
  {Teo}}]{Baryakhtar:2017ngi}%
  \BibitemOpen
  \bibfield  {author} {\bibinfo {author} {\bibfnamefont {M.}~\bibnamefont
  {Baryakhtar}}, \bibinfo {author} {\bibfnamefont {R.}~\bibnamefont {Lasenby}},
  \ and\ \bibinfo {author} {\bibfnamefont {M.}~\bibnamefont {Teo}},\ }\href
  {\doibase 10.1103/PhysRevD.96.035019} {\bibfield  {journal} {\bibinfo
  {journal} {Phys. Rev. D}\ }\textbf {\bibinfo {volume} {96}},\ \bibinfo
  {pages} {035019} (\bibinfo {year} {2017})},\ \Eprint
  {http://arxiv.org/abs/1704.05081} {arXiv:1704.05081 [hep-ph]} \BibitemShut
  {NoStop}%
\bibitem [{\citenamefont {Cardoso}\ \emph {et~al.}(2018)\citenamefont
  {Cardoso}, \citenamefont {Dias}, \citenamefont {Hartnett}, \citenamefont
  {Middleton}, \citenamefont {Pani},\ and\ \citenamefont
  {Santos}}]{Cardoso:2018tly}%
  \BibitemOpen
  \bibfield  {author} {\bibinfo {author} {\bibfnamefont {V.}~\bibnamefont
  {Cardoso}}, \bibinfo {author} {\bibfnamefont {O.~J.~C.}\ \bibnamefont
  {Dias}}, \bibinfo {author} {\bibfnamefont {G.~S.}\ \bibnamefont {Hartnett}},
  \bibinfo {author} {\bibfnamefont {M.}~\bibnamefont {Middleton}}, \bibinfo
  {author} {\bibfnamefont {P.}~\bibnamefont {Pani}}, \ and\ \bibinfo {author}
  {\bibfnamefont {J.~E.}\ \bibnamefont {Santos}},\ }\href {\doibase
  10.1088/1475-7516/2018/03/043} {\bibfield  {journal} {\bibinfo  {journal}
  {JCAP}\ }\textbf {\bibinfo {volume} {03}},\ \bibinfo {pages} {043} (\bibinfo
  {year} {2018})},\ \Eprint {http://arxiv.org/abs/1801.01420} {arXiv:1801.01420
  [gr-qc]} \BibitemShut {NoStop}%
\bibitem [{\citenamefont {Arvanitaki}\ \emph {et~al.}(2017)\citenamefont
  {Arvanitaki}, \citenamefont {Baryakhtar}, \citenamefont {Dimopoulos},
  \citenamefont {Dubovsky},\ and\ \citenamefont {Lasenby}}]{Arvanitaki2017}%
  \BibitemOpen
  \bibfield  {author} {\bibinfo {author} {\bibfnamefont {A.}~\bibnamefont
  {Arvanitaki}}, \bibinfo {author} {\bibfnamefont {M.}~\bibnamefont
  {Baryakhtar}}, \bibinfo {author} {\bibfnamefont {S.}~\bibnamefont
  {Dimopoulos}}, \bibinfo {author} {\bibfnamefont {S.}~\bibnamefont
  {Dubovsky}}, \ and\ \bibinfo {author} {\bibfnamefont {R.}~\bibnamefont
  {Lasenby}},\ }\href {\doibase 10.1103/PhysRevD.95.043001} {\bibfield
  {journal} {\bibinfo  {journal} {Phys. Rev. D}\ }\textbf {\bibinfo {volume}
  {95}},\ \bibinfo {pages} {043001} (\bibinfo {year} {2017})}\BibitemShut
  {NoStop}%
\bibitem [{\citenamefont {Ng}\ \emph {et~al.}(2021{\natexlab{a}})\citenamefont
  {Ng}, \citenamefont {Vitale}, \citenamefont {Hannuksela},\ and\ \citenamefont
  {Li}}]{Ng:2020ruv}%
  \BibitemOpen
  \bibfield  {author} {\bibinfo {author} {\bibfnamefont {K.~K.~Y.}\
  \bibnamefont {Ng}}, \bibinfo {author} {\bibfnamefont {S.}~\bibnamefont
  {Vitale}}, \bibinfo {author} {\bibfnamefont {O.~A.}\ \bibnamefont
  {Hannuksela}}, \ and\ \bibinfo {author} {\bibfnamefont {T.~G.~F.}\
  \bibnamefont {Li}},\ }\href {\doibase 10.1103/PhysRevLett.126.151102}
  {\bibfield  {journal} {\bibinfo  {journal} {Phys. Rev. Lett.}\ }\textbf
  {\bibinfo {volume} {126}},\ \bibinfo {pages} {151102} (\bibinfo {year}
  {2021}{\natexlab{a}})},\ \Eprint {http://arxiv.org/abs/2011.06010}
  {arXiv:2011.06010 [gr-qc]} \BibitemShut {NoStop}%
\bibitem [{\citenamefont {Ng}\ \emph {et~al.}(2021{\natexlab{b}})\citenamefont
  {Ng}, \citenamefont {Hannuksela}, \citenamefont {Vitale},\ and\ \citenamefont
  {Li}}]{Ng:2019jsx}%
  \BibitemOpen
  \bibfield  {author} {\bibinfo {author} {\bibfnamefont {K.~K.~Y.}\
  \bibnamefont {Ng}}, \bibinfo {author} {\bibfnamefont {O.~A.}\ \bibnamefont
  {Hannuksela}}, \bibinfo {author} {\bibfnamefont {S.}~\bibnamefont {Vitale}},
  \ and\ \bibinfo {author} {\bibfnamefont {T.~G.~F.}\ \bibnamefont {Li}},\
  }\href {\doibase 10.1103/PhysRevD.103.063010} {\bibfield  {journal} {\bibinfo
   {journal} {Phys. Rev. D}\ }\textbf {\bibinfo {volume} {103}},\ \bibinfo
  {pages} {063010} (\bibinfo {year} {2021}{\natexlab{b}})},\ \Eprint
  {http://arxiv.org/abs/1908.02312} {arXiv:1908.02312 [gr-qc]} \BibitemShut
  {NoStop}%
\bibitem [{\citenamefont {Ghosh}\ and\ \citenamefont
  {Sachdeva}(2021)}]{Ghosh:2021zuf}%
  \BibitemOpen
  \bibfield  {author} {\bibinfo {author} {\bibfnamefont {D.}~\bibnamefont
  {Ghosh}}\ and\ \bibinfo {author} {\bibfnamefont {D.}~\bibnamefont
  {Sachdeva}},\ }\href {\doibase 10.1103/PhysRevD.103.095028} {\bibfield
  {journal} {\bibinfo  {journal} {Phys. Rev. D}\ }\textbf {\bibinfo {volume}
  {103}},\ \bibinfo {pages} {095028} (\bibinfo {year} {2021})},\ \Eprint
  {http://arxiv.org/abs/2102.08857} {arXiv:2102.08857 [astro-ph.HE]}
  \BibitemShut {NoStop}%
\bibitem [{\citenamefont {Brito}\ \emph {et~al.}(2017)\citenamefont {Brito},
  \citenamefont {Ghosh}, \citenamefont {Barausse}, \citenamefont {Berti},
  \citenamefont {Cardoso}, \citenamefont {Dvorkin}, \citenamefont {Klein},\
  and\ \citenamefont {Pani}}]{Brito:2017zvb}%
  \BibitemOpen
  \bibfield  {author} {\bibinfo {author} {\bibfnamefont {R.}~\bibnamefont
  {Brito}}, \bibinfo {author} {\bibfnamefont {S.}~\bibnamefont {Ghosh}},
  \bibinfo {author} {\bibfnamefont {E.}~\bibnamefont {Barausse}}, \bibinfo
  {author} {\bibfnamefont {E.}~\bibnamefont {Berti}}, \bibinfo {author}
  {\bibfnamefont {V.}~\bibnamefont {Cardoso}}, \bibinfo {author} {\bibfnamefont
  {I.}~\bibnamefont {Dvorkin}}, \bibinfo {author} {\bibfnamefont
  {A.}~\bibnamefont {Klein}}, \ and\ \bibinfo {author} {\bibfnamefont
  {P.}~\bibnamefont {Pani}},\ }\href {\doibase 10.1103/PhysRevD.96.064050}
  {\bibfield  {journal} {\bibinfo  {journal} {Phys. Rev. D}\ }\textbf {\bibinfo
  {volume} {96}},\ \bibinfo {pages} {064050} (\bibinfo {year} {2017})},\
  \Eprint {http://arxiv.org/abs/1706.06311} {arXiv:1706.06311 [gr-qc]}
  \BibitemShut {NoStop}%
\bibitem [{\citenamefont {Isi}\ \emph {et~al.}(2019)\citenamefont {Isi},
  \citenamefont {Sun}, \citenamefont {Brito},\ and\ \citenamefont
  {Melatos}}]{Isi2019}%
  \BibitemOpen
  \bibfield  {author} {\bibinfo {author} {\bibfnamefont {M.}~\bibnamefont
  {Isi}}, \bibinfo {author} {\bibfnamefont {L.}~\bibnamefont {Sun}}, \bibinfo
  {author} {\bibfnamefont {R.}~\bibnamefont {Brito}}, \ and\ \bibinfo {author}
  {\bibfnamefont {A.}~\bibnamefont {Melatos}},\ }\href {\doibase
  10.1103/PhysRevD.99.084042} {\bibfield  {journal} {\bibinfo  {journal} {Phys.
  Rev. D}\ }\textbf {\bibinfo {volume} {99}},\ \bibinfo {pages} {084042}
  (\bibinfo {year} {2019})}\BibitemShut {NoStop}%
\bibitem [{\citenamefont {Jones}\ \emph {et~al.}(2023)\citenamefont {Jones},
  \citenamefont {Sun}, \citenamefont {Siemonsen}, \citenamefont {East},
  \citenamefont {Scott},\ and\ \citenamefont {Wette}}]{Jones2023}%
  \BibitemOpen
  \bibfield  {author} {\bibinfo {author} {\bibfnamefont {D.}~\bibnamefont
  {Jones}}, \bibinfo {author} {\bibfnamefont {L.}~\bibnamefont {Sun}}, \bibinfo
  {author} {\bibfnamefont {N.}~\bibnamefont {Siemonsen}}, \bibinfo {author}
  {\bibfnamefont {W.~E.}\ \bibnamefont {East}}, \bibinfo {author}
  {\bibfnamefont {S.~M.}\ \bibnamefont {Scott}}, \ and\ \bibinfo {author}
  {\bibfnamefont {K.}~\bibnamefont {Wette}},\ }\href {\doibase
  10.1103/PhysRevD.108.064001} {\bibfield  {journal} {\bibinfo  {journal}
  {Phys. Rev. D}\ }\textbf {\bibinfo {volume} {108}},\ \bibinfo {pages}
  {064001} (\bibinfo {year} {2023})}\BibitemShut {NoStop}%
\bibitem [{\citenamefont {Tsukada}\ \emph {et~al.}(2019)\citenamefont
  {Tsukada}, \citenamefont {Callister}, \citenamefont {Matas},\ and\
  \citenamefont {Meyers}}]{Tsukada:2018mbp}%
  \BibitemOpen
  \bibfield  {author} {\bibinfo {author} {\bibfnamefont {L.}~\bibnamefont
  {Tsukada}}, \bibinfo {author} {\bibfnamefont {T.}~\bibnamefont {Callister}},
  \bibinfo {author} {\bibfnamefont {A.}~\bibnamefont {Matas}}, \ and\ \bibinfo
  {author} {\bibfnamefont {P.}~\bibnamefont {Meyers}},\ }\href {\doibase
  10.1103/PhysRevD.99.103015} {\bibfield  {journal} {\bibinfo  {journal} {Phys.
  Rev. D}\ }\textbf {\bibinfo {volume} {99}},\ \bibinfo {pages} {103015}
  (\bibinfo {year} {2019})},\ \Eprint {http://arxiv.org/abs/1812.09622}
  {arXiv:1812.09622 [astro-ph.HE]} \BibitemShut {NoStop}%
\bibitem [{\citenamefont {Tsukada}\ \emph {et~al.}(2021)\citenamefont
  {Tsukada}, \citenamefont {Brito}, \citenamefont {East},\ and\ \citenamefont
  {Siemonsen}}]{Tsukada:2020lgt}%
  \BibitemOpen
  \bibfield  {author} {\bibinfo {author} {\bibfnamefont {L.}~\bibnamefont
  {Tsukada}}, \bibinfo {author} {\bibfnamefont {R.}~\bibnamefont {Brito}},
  \bibinfo {author} {\bibfnamefont {W.~E.}\ \bibnamefont {East}}, \ and\
  \bibinfo {author} {\bibfnamefont {N.}~\bibnamefont {Siemonsen}},\ }\href
  {\doibase 10.1103/PhysRevD.103.083005} {\bibfield  {journal} {\bibinfo
  {journal} {Phys. Rev. D}\ }\textbf {\bibinfo {volume} {103}},\ \bibinfo
  {pages} {083005} (\bibinfo {year} {2021})},\ \Eprint
  {http://arxiv.org/abs/2011.06995} {arXiv:2011.06995 [astro-ph.HE]}
  \BibitemShut {NoStop}%
\bibitem [{\citenamefont {Abbott}\ \emph
  {et~al.}(2022{\natexlab{a}})\citenamefont {Abbott} \emph
  {et~al.}}]{O3_all-sky_scalar_bosons}%
  \BibitemOpen
  \bibfield  {author} {\bibinfo {author} {\bibfnamefont {R.}~\bibnamefont
  {Abbott}} \emph {et~al.} (\bibinfo {collaboration} {The LIGO Scientific
  Collaboration, the Virgo Collaboration, and the KAGRA Collaboration}),\
  }\href {\doibase 10.1103/PhysRevD.105.102001} {\bibfield  {journal} {\bibinfo
   {journal} {Phys. Rev. D}\ }\textbf {\bibinfo {volume} {105}},\ \bibinfo
  {pages} {102001} (\bibinfo {year} {2022}{\natexlab{a}})}\BibitemShut
  {NoStop}%
\bibitem [{\citenamefont {Palomba}\ \emph {et~al.}(2019)\citenamefont {Palomba}
  \emph {et~al.}}]{Palomba2019}%
  \BibitemOpen
  \bibfield  {author} {\bibinfo {author} {\bibfnamefont {C.}~\bibnamefont
  {Palomba}} \emph {et~al.},\ }\href {\doibase 10.1103/PhysRevLett.123.171101}
  {\bibfield  {journal} {\bibinfo  {journal} {Phys. Rev. Lett.}\ }\textbf
  {\bibinfo {volume} {123}},\ \bibinfo {pages} {171101} (\bibinfo {year}
  {2019})}\BibitemShut {NoStop}%
\bibitem [{\citenamefont {Dergachev}\ and\ \citenamefont
  {Papa}(2019)}]{Dergachev2019}%
  \BibitemOpen
  \bibfield  {author} {\bibinfo {author} {\bibfnamefont {V.}~\bibnamefont
  {Dergachev}}\ and\ \bibinfo {author} {\bibfnamefont {M.~A.}\ \bibnamefont
  {Papa}},\ }\href {\doibase 10.1103/PhysRevLett.123.101101} {\bibfield
  {journal} {\bibinfo  {journal} {Phys. Rev. Lett.}\ }\textbf {\bibinfo
  {volume} {123}},\ \bibinfo {pages} {101101} (\bibinfo {year}
  {2019})}\BibitemShut {NoStop}%
\bibitem [{\citenamefont {Abbott}\ \emph
  {et~al.}(2022{\natexlab{b}})\citenamefont {Abbott} \emph
  {et~al.}}]{O3_CWs_Milky_Way}%
  \BibitemOpen
  \bibfield  {author} {\bibinfo {author} {\bibfnamefont {R.}~\bibnamefont
  {Abbott}} \emph {et~al.} (\bibinfo {collaboration} {KAGRA, LIGO Scientific,
  VIRGO}),\ }\href {\doibase 10.1103/PhysRevD.106.042003} {\bibfield  {journal}
  {\bibinfo  {journal} {Phys. Rev. D}\ }\textbf {\bibinfo {volume} {106}},\
  \bibinfo {pages} {042003} (\bibinfo {year} {2022}{\natexlab{b}})}\BibitemShut
  {NoStop}%
\bibitem [{\citenamefont {Zhu}\ \emph {et~al.}(2020)\citenamefont {Zhu},
  \citenamefont {Baryakhtar}, \citenamefont {Papa}, \citenamefont {Tsuna},
  \citenamefont {Kawanaka},\ and\ \citenamefont {Eggenstein}}]{Zhu2020}%
  \BibitemOpen
  \bibfield  {author} {\bibinfo {author} {\bibfnamefont {S.~J.}\ \bibnamefont
  {Zhu}}, \bibinfo {author} {\bibfnamefont {M.}~\bibnamefont {Baryakhtar}},
  \bibinfo {author} {\bibfnamefont {M.~A.}\ \bibnamefont {Papa}}, \bibinfo
  {author} {\bibfnamefont {D.}~\bibnamefont {Tsuna}}, \bibinfo {author}
  {\bibfnamefont {N.}~\bibnamefont {Kawanaka}}, \ and\ \bibinfo {author}
  {\bibfnamefont {H.-B.}\ \bibnamefont {Eggenstein}},\ }\href {\doibase
  10.1103/PhysRevD.102.063020} {\bibfield  {journal} {\bibinfo  {journal}
  {Phys. Rev. D}\ }\textbf {\bibinfo {volume} {102}},\ \bibinfo {pages}
  {063020} (\bibinfo {year} {2020})}\BibitemShut {NoStop}%
\bibitem [{\citenamefont {Sun}\ \emph {et~al.}(2020{\natexlab{a}})\citenamefont
  {Sun}, \citenamefont {Brito},\ and\ \citenamefont
  {Isi}}]{O2_CygnusX1_scalar_bosons}%
  \BibitemOpen
  \bibfield  {author} {\bibinfo {author} {\bibfnamefont {L.}~\bibnamefont
  {Sun}}, \bibinfo {author} {\bibfnamefont {R.}~\bibnamefont {Brito}}, \ and\
  \bibinfo {author} {\bibfnamefont {M.}~\bibnamefont {Isi}},\ }\href {\doibase
  10.1103/PhysRevD.101.063020} {\bibfield  {journal} {\bibinfo  {journal}
  {Phys. Rev. D}\ }\textbf {\bibinfo {volume} {101}},\ \bibinfo {pages}
  {063020} (\bibinfo {year} {2020}{\natexlab{a}})}\BibitemShut {NoStop}%
\bibitem [{\citenamefont {Collaviti}\ \emph {et~al.}(2024)\citenamefont
  {Collaviti}, \citenamefont {Sun}, \citenamefont {Galanis},\ and\
  \citenamefont {Baryakhtar}}]{Collaviti2024}%
  \BibitemOpen
  \bibfield  {author} {\bibinfo {author} {\bibfnamefont {S.}~\bibnamefont
  {Collaviti}}, \bibinfo {author} {\bibfnamefont {L.}~\bibnamefont {Sun}},
  \bibinfo {author} {\bibfnamefont {M.}~\bibnamefont {Galanis}}, \ and\
  \bibinfo {author} {\bibfnamefont {M.}~\bibnamefont {Baryakhtar}},\ }\href
  {https://arxiv.org/abs/2407.04304} {\  (\bibinfo {year} {2024})},\ \Eprint
  {http://arxiv.org/abs/2407.04304} {arXiv:2407.04304 [gr-qc]} \BibitemShut
  {NoStop}%
\bibitem [{\citenamefont {Aasi}\ \emph {et~al.}(2015)\citenamefont {Aasi} \emph
  {et~al.}}]{LIGOScientific:2014pky}%
  \BibitemOpen
  \bibfield  {author} {\bibinfo {author} {\bibfnamefont {J.}~\bibnamefont
  {Aasi}} \emph {et~al.} (\bibinfo {collaboration} {LIGO Scientific}),\ }\href
  {\doibase 10.1088/0264-9381/32/7/074001} {\bibfield  {journal} {\bibinfo
  {journal} {Class. Quant. Grav.}\ }\textbf {\bibinfo {volume} {32}},\ \bibinfo
  {pages} {074001} (\bibinfo {year} {2015})},\ \Eprint
  {http://arxiv.org/abs/1411.4547} {arXiv:1411.4547 [gr-qc]} \BibitemShut
  {NoStop}%
\bibitem [{\citenamefont {Acernese}\ \emph {et~al.}(2015)\citenamefont
  {Acernese} \emph {et~al.}}]{VIRGO:2014yos}%
  \BibitemOpen
  \bibfield  {author} {\bibinfo {author} {\bibfnamefont {F.}~\bibnamefont
  {Acernese}} \emph {et~al.} (\bibinfo {collaboration} {VIRGO}),\ }\href
  {\doibase 10.1088/0264-9381/32/2/024001} {\bibfield  {journal} {\bibinfo
  {journal} {Class. Quant. Grav.}\ }\textbf {\bibinfo {volume} {32}},\ \bibinfo
  {pages} {024001} (\bibinfo {year} {2015})},\ \Eprint
  {http://arxiv.org/abs/1408.3978} {arXiv:1408.3978 [gr-qc]} \BibitemShut
  {NoStop}%
\bibitem [{\citenamefont {Akutsu}\ \emph {et~al.}(2021)\citenamefont {Akutsu}
  \emph {et~al.}}]{KAGRA:2020tym}%
  \BibitemOpen
  \bibfield  {author} {\bibinfo {author} {\bibfnamefont {T.}~\bibnamefont
  {Akutsu}} \emph {et~al.} (\bibinfo {collaboration} {KAGRA}),\ }\href
  {\doibase 10.1093/ptep/ptaa125} {\bibfield  {journal} {\bibinfo  {journal}
  {PTEP}\ }\textbf {\bibinfo {volume} {2021}},\ \bibinfo {pages} {05A101}
  (\bibinfo {year} {2021})},\ \Eprint {http://arxiv.org/abs/2005.05574}
  {arXiv:2005.05574 [physics.ins-det]} \BibitemShut {NoStop}%
\bibitem [{\citenamefont {Abac}\ \emph
  {et~al.}(2025{\natexlab{a}})\citenamefont {Abac} \emph {et~al.}}]{GW231123}%
  \BibitemOpen
  \bibfield  {author} {\bibinfo {author} {\bibfnamefont {A.~G.}\ \bibnamefont
  {Abac}} \emph {et~al.} (\bibinfo {collaboration} {The LIGO Scientific
  Collaboration, the Virgo Collaboration, and the KAGRA Collaboration}),\
  }\href {https://arxiv.org/abs/2507.08219} {\  (\bibinfo {year}
  {2025}{\natexlab{a}})},\ \Eprint {http://arxiv.org/abs/2507.08219}
  {arXiv:2507.08219 [astro-ph.HE]} \BibitemShut {NoStop}%
\bibitem [{\citenamefont {{LIGO Scientific, Virgo, and KAGRA
  Collaboration}}(2025)}]{zenodo_GW231123}%
  \BibitemOpen
  \bibfield  {author} {\bibinfo {author} {\bibnamefont {{LIGO Scientific,
  Virgo, and KAGRA Collaboration}}},\ }\href {\doibase 10.5281/zenodo.16004263}
  {\enquote {\bibinfo {title} {{GW231123: a Binary Black Hole Merger with Total
  Mass 190-265\,$M_\odot$ --- Data Release}},}\ } (\bibinfo {year}
  {2025})\BibitemShut {NoStop}%
\bibitem [{\citenamefont {Abbott}\ \emph {et~al.}(2024)\citenamefont {Abbott}
  \emph {et~al.}}]{GWTC-2.1}%
  \BibitemOpen
  \bibfield  {author} {\bibinfo {author} {\bibfnamefont {R.}~\bibnamefont
  {Abbott}} \emph {et~al.} (\bibinfo {collaboration} {The LIGO Scientific
  Collaboration and the Virgo Collaboration}),\ }\href {\doibase
  10.1103/PhysRevD.109.022001} {\bibfield  {journal} {\bibinfo  {journal}
  {Phys. Rev. D}\ }\textbf {\bibinfo {volume} {109}},\ \bibinfo {pages}
  {022001} (\bibinfo {year} {2024})}\BibitemShut {NoStop}%
\bibitem [{\citenamefont {{LIGO Scientific Collaboration and Virgo
  Collaboration}}(2022)}]{zenodo_GW190517}%
  \BibitemOpen
  \bibfield  {author} {\bibinfo {author} {\bibnamefont {{LIGO Scientific
  Collaboration and Virgo Collaboration}}},\ }\href {\doibase
  10.5281/zenodo.6513631} {\enquote {\bibinfo {title} {{GWTC-2.1: Deep Extended
  Catalog of Compact Binary Coalescences Observed by LIGO and Virgo During the
  First Half of the Third Observing Run - Parameter Estimation Data
  Release}},}\ } (\bibinfo {year} {2022})\BibitemShut {NoStop}%
\bibitem [{\citenamefont {Abac}\ \emph
  {et~al.}(2025{\natexlab{b}})\citenamefont {Abac} \emph
  {et~al.}}]{LIGOScientific:2025slb}%
  \BibitemOpen
  \bibfield  {author} {\bibinfo {author} {\bibfnamefont {A.~G.}\ \bibnamefont
  {Abac}} \emph {et~al.} (\bibinfo {collaboration} {LIGO Scientific, VIRGO,
  KAGRA}),\ }\href@noop {} {\  (\bibinfo {year} {2025}{\natexlab{b}})},\
  \Eprint {http://arxiv.org/abs/2508.18082} {arXiv:2508.18082 [gr-qc]}
  \BibitemShut {NoStop}%
\bibitem [{\citenamefont {Ray}\ \emph {et~al.}(2025)\citenamefont {Ray},
  \citenamefont {Banagiri}, \citenamefont {Thrane},\ and\ \citenamefont
  {Lasky}}]{Ray:2025rtt}%
  \BibitemOpen
  \bibfield  {author} {\bibinfo {author} {\bibfnamefont {A.}~\bibnamefont
  {Ray}}, \bibinfo {author} {\bibfnamefont {S.}~\bibnamefont {Banagiri}},
  \bibinfo {author} {\bibfnamefont {E.}~\bibnamefont {Thrane}}, \ and\ \bibinfo
  {author} {\bibfnamefont {P.~D.}\ \bibnamefont {Lasky}},\ }\href@noop {} {\
  (\bibinfo {year} {2025})},\ \Eprint {http://arxiv.org/abs/2510.07228}
  {arXiv:2510.07228 [gr-qc]} \BibitemShut {NoStop}%
\bibitem [{\citenamefont {Ashton}\ \emph {et~al.}(2022)\citenamefont {Ashton},
  \citenamefont {Thiele}, \citenamefont {Lecoeuche}, \citenamefont {McIver},\
  and\ \citenamefont {Nuttall}}]{Ashton:2021tvz}%
  \BibitemOpen
  \bibfield  {author} {\bibinfo {author} {\bibfnamefont {G.}~\bibnamefont
  {Ashton}}, \bibinfo {author} {\bibfnamefont {S.}~\bibnamefont {Thiele}},
  \bibinfo {author} {\bibfnamefont {Y.}~\bibnamefont {Lecoeuche}}, \bibinfo
  {author} {\bibfnamefont {J.}~\bibnamefont {McIver}}, \ and\ \bibinfo {author}
  {\bibfnamefont {L.~K.}\ \bibnamefont {Nuttall}},\ }\href {\doibase
  10.1088/1361-6382/ac8094} {\bibfield  {journal} {\bibinfo  {journal} {Class.
  Quant. Grav.}\ }\textbf {\bibinfo {volume} {39}},\ \bibinfo {pages} {175004}
  (\bibinfo {year} {2022})},\ \Eprint {http://arxiv.org/abs/2110.02689}
  {arXiv:2110.02689 [gr-qc]} \BibitemShut {NoStop}%
\bibitem [{\citenamefont {Siemonsen}\ \emph
  {et~al.}(2023{\natexlab{a}})\citenamefont {Siemonsen}, \citenamefont {May},\
  and\ \citenamefont {East}}]{Siemonsen:2022yyf}%
  \BibitemOpen
  \bibfield  {author} {\bibinfo {author} {\bibfnamefont {N.}~\bibnamefont
  {Siemonsen}}, \bibinfo {author} {\bibfnamefont {T.}~\bibnamefont {May}}, \
  and\ \bibinfo {author} {\bibfnamefont {W.~E.}\ \bibnamefont {East}},\ }\href
  {\doibase 10.1103/PhysRevD.107.104003} {\bibfield  {journal} {\bibinfo
  {journal} {Phys. Rev. D}\ }\textbf {\bibinfo {volume} {107}},\ \bibinfo
  {pages} {104003} (\bibinfo {year} {2023}{\natexlab{a}})},\ \Eprint
  {http://arxiv.org/abs/2211.03845} {arXiv:2211.03845 [gr-qc]} \BibitemShut
  {NoStop}%
\bibitem [{\citenamefont {May}\ \emph {et~al.}(2025)\citenamefont {May},
  \citenamefont {East},\ and\ \citenamefont {Siemonsen}}]{May2025}%
  \BibitemOpen
  \bibfield  {author} {\bibinfo {author} {\bibfnamefont {T.}~\bibnamefont
  {May}}, \bibinfo {author} {\bibfnamefont {W.~E.}\ \bibnamefont {East}}, \
  and\ \bibinfo {author} {\bibfnamefont {N.}~\bibnamefont {Siemonsen}},\ }\href
  {\doibase 10.1103/PhysRevD.111.044062} {\bibfield  {journal} {\bibinfo
  {journal} {Phys. Rev. D}\ }\textbf {\bibinfo {volume} {111}},\ \bibinfo
  {pages} {044062} (\bibinfo {year} {2025})}\BibitemShut {NoStop}%
\bibitem [{\citenamefont {Omiya}\ \emph {et~al.}(2022)\citenamefont {Omiya},
  \citenamefont {Takahashi},\ and\ \citenamefont {Tanaka}}]{Omiya:2022mwv}%
  \BibitemOpen
  \bibfield  {author} {\bibinfo {author} {\bibfnamefont {H.}~\bibnamefont
  {Omiya}}, \bibinfo {author} {\bibfnamefont {T.}~\bibnamefont {Takahashi}}, \
  and\ \bibinfo {author} {\bibfnamefont {T.}~\bibnamefont {Tanaka}},\ }\href
  {\doibase 10.1093/ptep/ptac058} {\bibfield  {journal} {\bibinfo  {journal}
  {PTEP}\ }\textbf {\bibinfo {volume} {2022}},\ \bibinfo {pages} {043E03}
  (\bibinfo {year} {2022})},\ \Eprint {http://arxiv.org/abs/2201.04382}
  {arXiv:2201.04382 [gr-qc]} \BibitemShut {NoStop}%
\bibitem [{\citenamefont {Witte}\ and\ \citenamefont
  {Mummery}(2025)}]{Witte:2024drg}%
  \BibitemOpen
  \bibfield  {author} {\bibinfo {author} {\bibfnamefont {S.~J.}\ \bibnamefont
  {Witte}}\ and\ \bibinfo {author} {\bibfnamefont {A.}~\bibnamefont
  {Mummery}},\ }\href {\doibase 10.1103/PhysRevD.111.083044} {\bibfield
  {journal} {\bibinfo  {journal} {Phys. Rev. D}\ }\textbf {\bibinfo {volume}
  {111}},\ \bibinfo {pages} {083044} (\bibinfo {year} {2025})},\ \Eprint
  {http://arxiv.org/abs/2412.03655} {arXiv:2412.03655 [hep-ph]} \BibitemShut
  {NoStop}%
\bibitem [{\citenamefont {Siemonsen}\ \emph
  {et~al.}(2023{\natexlab{b}})\citenamefont {Siemonsen}, \citenamefont
  {Mondino}, \citenamefont {Egana-Ugrinovic}, \citenamefont {Huang},
  \citenamefont {Baryakhtar},\ and\ \citenamefont {East}}]{Siemonsen:2022ivj}%
  \BibitemOpen
  \bibfield  {author} {\bibinfo {author} {\bibfnamefont {N.}~\bibnamefont
  {Siemonsen}}, \bibinfo {author} {\bibfnamefont {C.}~\bibnamefont {Mondino}},
  \bibinfo {author} {\bibfnamefont {D.}~\bibnamefont {Egana-Ugrinovic}},
  \bibinfo {author} {\bibfnamefont {J.}~\bibnamefont {Huang}}, \bibinfo
  {author} {\bibfnamefont {M.}~\bibnamefont {Baryakhtar}}, \ and\ \bibinfo
  {author} {\bibfnamefont {W.~E.}\ \bibnamefont {East}},\ }\href {\doibase
  10.1103/PhysRevD.107.075025} {\bibfield  {journal} {\bibinfo  {journal}
  {Phys. Rev. D}\ }\textbf {\bibinfo {volume} {107}},\ \bibinfo {pages}
  {075025} (\bibinfo {year} {2023}{\natexlab{b}})},\ \Eprint
  {http://arxiv.org/abs/2212.09772} {arXiv:2212.09772 [astro-ph.HE]}
  \BibitemShut {NoStop}%
\bibitem [{\citenamefont {Xin}\ and\ \citenamefont {Most}(2025)}]{Xin:2024trp}%
  \BibitemOpen
  \bibfield  {author} {\bibinfo {author} {\bibfnamefont {S.}~\bibnamefont
  {Xin}}\ and\ \bibinfo {author} {\bibfnamefont {E.~R.}\ \bibnamefont {Most}},\
  }\href {\doibase 10.1103/PhysRevD.111.063050} {\bibfield  {journal} {\bibinfo
   {journal} {Phys. Rev. D}\ }\textbf {\bibinfo {volume} {111}},\ \bibinfo
  {pages} {063050} (\bibinfo {year} {2025})},\ \Eprint
  {http://arxiv.org/abs/2406.02992} {arXiv:2406.02992 [astro-ph.HE]}
  \BibitemShut {NoStop}%
\bibitem [{\citenamefont {Fukuda}\ and\ \citenamefont
  {Nakayama}(2020)}]{Fukuda:2019ewf}%
  \BibitemOpen
  \bibfield  {author} {\bibinfo {author} {\bibfnamefont {H.}~\bibnamefont
  {Fukuda}}\ and\ \bibinfo {author} {\bibfnamefont {K.}~\bibnamefont
  {Nakayama}},\ }\href {\doibase 10.1007/JHEP01(2020)128} {\bibfield  {journal}
  {\bibinfo  {journal} {JHEP}\ }\textbf {\bibinfo {volume} {01}},\ \bibinfo
  {pages} {128} (\bibinfo {year} {2020})},\ \Eprint
  {http://arxiv.org/abs/1910.06308} {arXiv:1910.06308 [hep-ph]} \BibitemShut
  {NoStop}%
\bibitem [{\citenamefont {Jones}\ \emph {et~al.}(2025)\citenamefont {Jones},
  \citenamefont {Siemonsen}, \citenamefont {Sun}, \citenamefont {East},
  \citenamefont {Miller}, \citenamefont {Wette},\ and\ \citenamefont
  {Piccinni}}]{Jones:2024fpg}%
  \BibitemOpen
  \bibfield  {author} {\bibinfo {author} {\bibfnamefont {D.}~\bibnamefont
  {Jones}}, \bibinfo {author} {\bibfnamefont {N.}~\bibnamefont {Siemonsen}},
  \bibinfo {author} {\bibfnamefont {L.}~\bibnamefont {Sun}}, \bibinfo {author}
  {\bibfnamefont {W.~E.}\ \bibnamefont {East}}, \bibinfo {author}
  {\bibfnamefont {A.~L.}\ \bibnamefont {Miller}}, \bibinfo {author}
  {\bibfnamefont {K.}~\bibnamefont {Wette}}, \ and\ \bibinfo {author}
  {\bibfnamefont {O.~J.}\ \bibnamefont {Piccinni}},\ }\href {\doibase
  10.1103/PhysRevD.111.063028} {\bibfield  {journal} {\bibinfo  {journal}
  {Phys. Rev. D}\ }\textbf {\bibinfo {volume} {111}},\ \bibinfo {pages}
  {063028} (\bibinfo {year} {2025})},\ \Eprint
  {http://arxiv.org/abs/2412.00320} {arXiv:2412.00320 [gr-qc]} \BibitemShut
  {NoStop}%
\bibitem [{\citenamefont {East}(2022)}]{East:2022ppo}%
  \BibitemOpen
  \bibfield  {author} {\bibinfo {author} {\bibfnamefont {W.~E.}\ \bibnamefont
  {East}},\ }\href {\doibase 10.1103/PhysRevLett.129.141103} {\bibfield
  {journal} {\bibinfo  {journal} {Phys. Rev. Lett.}\ }\textbf {\bibinfo
  {volume} {129}},\ \bibinfo {pages} {141103} (\bibinfo {year} {2022})},\
  \Eprint {http://arxiv.org/abs/2205.03417} {arXiv:2205.03417 [hep-ph]}
  \BibitemShut {NoStop}%
\bibitem [{\citenamefont {East}\ and\ \citenamefont
  {Huang}(2022)}]{East:2022rsi}%
  \BibitemOpen
  \bibfield  {author} {\bibinfo {author} {\bibfnamefont {W.~E.}\ \bibnamefont
  {East}}\ and\ \bibinfo {author} {\bibfnamefont {J.}~\bibnamefont {Huang}},\
  }\href {\doibase 10.1007/JHEP12(2022)089} {\bibfield  {journal} {\bibinfo
  {journal} {JHEP}\ }\textbf {\bibinfo {volume} {12}},\ \bibinfo {pages} {089}
  (\bibinfo {year} {2022})},\ \Eprint {http://arxiv.org/abs/2206.12432}
  {arXiv:2206.12432 [hep-ph]} \BibitemShut {NoStop}%
\bibitem [{\citenamefont {Brzeminski}\ \emph {et~al.}(2025)\citenamefont
  {Brzeminski}, \citenamefont {Hook}, \citenamefont {Huang},\ and\
  \citenamefont {Ristow}}]{Brzeminski:2024drp}%
  \BibitemOpen
  \bibfield  {author} {\bibinfo {author} {\bibfnamefont {D.}~\bibnamefont
  {Brzeminski}}, \bibinfo {author} {\bibfnamefont {A.}~\bibnamefont {Hook}},
  \bibinfo {author} {\bibfnamefont {J.}~\bibnamefont {Huang}}, \ and\ \bibinfo
  {author} {\bibfnamefont {C.}~\bibnamefont {Ristow}},\ }\href {\doibase
  10.1007/JHEP01(2025)007} {\bibfield  {journal} {\bibinfo  {journal} {JHEP}\
  }\textbf {\bibinfo {volume} {01}},\ \bibinfo {pages} {007} (\bibinfo {year}
  {2025})},\ \Eprint {http://arxiv.org/abs/2407.18991} {arXiv:2407.18991
  [hep-ph]} \BibitemShut {NoStop}%
\bibitem [{\citenamefont {Portegies~Zwart}\ and\ \citenamefont
  {McMillan}(2000)}]{PortegiesZwart:1999nm}%
  \BibitemOpen
  \bibfield  {author} {\bibinfo {author} {\bibfnamefont {S.~F.}\ \bibnamefont
  {Portegies~Zwart}}\ and\ \bibinfo {author} {\bibfnamefont {S.}~\bibnamefont
  {McMillan}},\ }\href {\doibase 10.1086/312422} {\bibfield  {journal}
  {\bibinfo  {journal} {Astrophys. J. Lett.}\ }\textbf {\bibinfo {volume}
  {528}},\ \bibinfo {pages} {L17} (\bibinfo {year} {2000})},\ \Eprint
  {http://arxiv.org/abs/astro-ph/9910061} {arXiv:astro-ph/9910061} \BibitemShut
  {NoStop}%
\bibitem [{\citenamefont {Dominik}\ \emph {et~al.}(2013)\citenamefont
  {Dominik}, \citenamefont {Belczynski}, \citenamefont {Fryer}, \citenamefont
  {Holz}, \citenamefont {Berti}, \citenamefont {Bulik}, \citenamefont
  {Mandel},\ and\ \citenamefont {O'Shaughnessy}}]{Dominik:2013tma}%
  \BibitemOpen
  \bibfield  {author} {\bibinfo {author} {\bibfnamefont {M.}~\bibnamefont
  {Dominik}}, \bibinfo {author} {\bibfnamefont {K.}~\bibnamefont {Belczynski}},
  \bibinfo {author} {\bibfnamefont {C.}~\bibnamefont {Fryer}}, \bibinfo
  {author} {\bibfnamefont {D.~E.}\ \bibnamefont {Holz}}, \bibinfo {author}
  {\bibfnamefont {E.}~\bibnamefont {Berti}}, \bibinfo {author} {\bibfnamefont
  {T.}~\bibnamefont {Bulik}}, \bibinfo {author} {\bibfnamefont
  {I.}~\bibnamefont {Mandel}}, \ and\ \bibinfo {author} {\bibfnamefont
  {R.}~\bibnamefont {O'Shaughnessy}},\ }\href {\doibase
  10.1088/0004-637X/779/1/72} {\bibfield  {journal} {\bibinfo  {journal}
  {Astrophys. J.}\ }\textbf {\bibinfo {volume} {779}},\ \bibinfo {pages} {72}
  (\bibinfo {year} {2013})},\ \Eprint {http://arxiv.org/abs/1308.1546}
  {arXiv:1308.1546 [astro-ph.HE]} \BibitemShut {NoStop}%
\bibitem [{\citenamefont {Bavera}\ \emph {et~al.}(2020)\citenamefont {Bavera},
  \citenamefont {Fragos}, \citenamefont {Qin}, \citenamefont {Zapartas},
  \citenamefont {Neijssel}, \citenamefont {Mandel}, \citenamefont {Batta},
  \citenamefont {Gaebel}, \citenamefont {Kimball},\ and\ \citenamefont
  {Stevenson}}]{Bavera:2020inc}%
  \BibitemOpen
  \bibfield  {author} {\bibinfo {author} {\bibfnamefont {S.~S.}\ \bibnamefont
  {Bavera}}, \bibinfo {author} {\bibfnamefont {T.}~\bibnamefont {Fragos}},
  \bibinfo {author} {\bibfnamefont {Y.}~\bibnamefont {Qin}}, \bibinfo {author}
  {\bibfnamefont {E.}~\bibnamefont {Zapartas}}, \bibinfo {author}
  {\bibfnamefont {C.~J.}\ \bibnamefont {Neijssel}}, \bibinfo {author}
  {\bibfnamefont {I.}~\bibnamefont {Mandel}}, \bibinfo {author} {\bibfnamefont
  {A.}~\bibnamefont {Batta}}, \bibinfo {author} {\bibfnamefont {S.~M.}\
  \bibnamefont {Gaebel}}, \bibinfo {author} {\bibfnamefont {C.}~\bibnamefont
  {Kimball}}, \ and\ \bibinfo {author} {\bibfnamefont {S.}~\bibnamefont
  {Stevenson}},\ }\href {\doibase 10.1051/0004-6361/201936204} {\bibfield
  {journal} {\bibinfo  {journal} {Astron. Astrophys.}\ }\textbf {\bibinfo
  {volume} {635}},\ \bibinfo {pages} {A97} (\bibinfo {year} {2020})},\ \Eprint
  {http://arxiv.org/abs/1906.12257} {arXiv:1906.12257 [astro-ph.HE]}
  \BibitemShut {NoStop}%
\bibitem [{\citenamefont {Rodriguez}\ \emph {et~al.}(2016)\citenamefont
  {Rodriguez}, \citenamefont {Chatterjee},\ and\ \citenamefont
  {Rasio}}]{Rodriguez:2016kxx}%
  \BibitemOpen
  \bibfield  {author} {\bibinfo {author} {\bibfnamefont {C.~L.}\ \bibnamefont
  {Rodriguez}}, \bibinfo {author} {\bibfnamefont {S.}~\bibnamefont
  {Chatterjee}}, \ and\ \bibinfo {author} {\bibfnamefont {F.~A.}\ \bibnamefont
  {Rasio}},\ }\href {\doibase 10.1103/PhysRevD.93.084029} {\bibfield  {journal}
  {\bibinfo  {journal} {Phys. Rev. D}\ }\textbf {\bibinfo {volume} {93}},\
  \bibinfo {pages} {084029} (\bibinfo {year} {2016})},\ \Eprint
  {http://arxiv.org/abs/1602.02444} {arXiv:1602.02444 [astro-ph.HE]}
  \BibitemShut {NoStop}%
\bibitem [{\citenamefont {Bartos}\ \emph {et~al.}(2017)\citenamefont {Bartos},
  \citenamefont {Kocsis}, \citenamefont {Haiman},\ and\ \citenamefont
  {M\'arka}}]{Bartos:2016dgn}%
  \BibitemOpen
  \bibfield  {author} {\bibinfo {author} {\bibfnamefont {I.}~\bibnamefont
  {Bartos}}, \bibinfo {author} {\bibfnamefont {B.}~\bibnamefont {Kocsis}},
  \bibinfo {author} {\bibfnamefont {Z.}~\bibnamefont {Haiman}}, \ and\ \bibinfo
  {author} {\bibfnamefont {S.}~\bibnamefont {M\'arka}},\ }\href {\doibase
  10.3847/1538-4357/835/2/165} {\bibfield  {journal} {\bibinfo  {journal}
  {Astrophys. J.}\ }\textbf {\bibinfo {volume} {835}},\ \bibinfo {pages} {165}
  (\bibinfo {year} {2017})},\ \Eprint {http://arxiv.org/abs/1602.03831}
  {arXiv:1602.03831 [astro-ph.HE]} \BibitemShut {NoStop}%
\bibitem [{\citenamefont {Yang}\ \emph {et~al.}(2019)\citenamefont {Yang} \emph
  {et~al.}}]{Yang:2019cbr}%
  \BibitemOpen
  \bibfield  {author} {\bibinfo {author} {\bibfnamefont {Y.}~\bibnamefont
  {Yang}} \emph {et~al.},\ }\href {\doibase 10.1103/PhysRevLett.123.181101}
  {\bibfield  {journal} {\bibinfo  {journal} {Phys. Rev. Lett.}\ }\textbf
  {\bibinfo {volume} {123}},\ \bibinfo {pages} {181101} (\bibinfo {year}
  {2019})},\ \Eprint {http://arxiv.org/abs/1906.09281} {arXiv:1906.09281
  [astro-ph.HE]} \BibitemShut {NoStop}%
\bibitem [{\citenamefont {Wilczek}(1978)}]{Wilczek:1977pj}%
  \BibitemOpen
  \bibfield  {author} {\bibinfo {author} {\bibfnamefont {F.}~\bibnamefont
  {Wilczek}},\ }\href {\doibase 10.1103/PhysRevLett.40.279} {\bibfield
  {journal} {\bibinfo  {journal} {Phys. Rev. Lett.}\ }\textbf {\bibinfo
  {volume} {40}},\ \bibinfo {pages} {279} (\bibinfo {year} {1978})}\BibitemShut
  {NoStop}%
\bibitem [{\citenamefont {Mirizzi}\ \emph {et~al.}(2009)\citenamefont
  {Mirizzi}, \citenamefont {Redondo},\ and\ \citenamefont
  {Sigl}}]{Mirizzi:2009iz}%
  \BibitemOpen
  \bibfield  {author} {\bibinfo {author} {\bibfnamefont {A.}~\bibnamefont
  {Mirizzi}}, \bibinfo {author} {\bibfnamefont {J.}~\bibnamefont {Redondo}}, \
  and\ \bibinfo {author} {\bibfnamefont {G.}~\bibnamefont {Sigl}},\ }\href
  {\doibase 10.1088/1475-7516/2009/03/026} {\bibfield  {journal} {\bibinfo
  {journal} {JCAP}\ }\textbf {\bibinfo {volume} {03}},\ \bibinfo {pages} {026}
  (\bibinfo {year} {2009})},\ \Eprint {http://arxiv.org/abs/0901.0014}
  {arXiv:0901.0014 [hep-ph]} \BibitemShut {NoStop}%
\bibitem [{\citenamefont {Caputo}\ \emph {et~al.}(2020)\citenamefont {Caputo},
  \citenamefont {Liu}, \citenamefont {Mishra-Sharma},\ and\ \citenamefont
  {Ruderman}}]{Caputo:2020bdy}%
  \BibitemOpen
  \bibfield  {author} {\bibinfo {author} {\bibfnamefont {A.}~\bibnamefont
  {Caputo}}, \bibinfo {author} {\bibfnamefont {H.}~\bibnamefont {Liu}},
  \bibinfo {author} {\bibfnamefont {S.}~\bibnamefont {Mishra-Sharma}}, \ and\
  \bibinfo {author} {\bibfnamefont {J.~T.}\ \bibnamefont {Ruderman}},\ }\href
  {\doibase 10.1103/PhysRevLett.125.221303} {\bibfield  {journal} {\bibinfo
  {journal} {Phys. Rev. Lett.}\ }\textbf {\bibinfo {volume} {125}},\ \bibinfo
  {pages} {221303} (\bibinfo {year} {2020})},\ \Eprint
  {http://arxiv.org/abs/2002.05165} {arXiv:2002.05165 [astro-ph.CO]}
  \BibitemShut {NoStop}%
\bibitem [{\citenamefont {Fixsen}\ \emph {et~al.}(1996)\citenamefont {Fixsen},
  \citenamefont {Cheng}, \citenamefont {Gales}, \citenamefont {Mather},
  \citenamefont {Shafer},\ and\ \citenamefont {Wright}}]{Fixsen:1996nj}%
  \BibitemOpen
  \bibfield  {author} {\bibinfo {author} {\bibfnamefont {D.~J.}\ \bibnamefont
  {Fixsen}}, \bibinfo {author} {\bibfnamefont {E.~S.}\ \bibnamefont {Cheng}},
  \bibinfo {author} {\bibfnamefont {J.~M.}\ \bibnamefont {Gales}}, \bibinfo
  {author} {\bibfnamefont {J.~C.}\ \bibnamefont {Mather}}, \bibinfo {author}
  {\bibfnamefont {R.~A.}\ \bibnamefont {Shafer}}, \ and\ \bibinfo {author}
  {\bibfnamefont {E.~L.}\ \bibnamefont {Wright}},\ }\href {\doibase
  10.1086/178173} {\bibfield  {journal} {\bibinfo  {journal} {Astrophys. J.}\
  }\textbf {\bibinfo {volume} {473}},\ \bibinfo {pages} {576} (\bibinfo {year}
  {1996})},\ \Eprint {http://arxiv.org/abs/astro-ph/9605054}
  {arXiv:astro-ph/9605054} \BibitemShut {NoStop}%
\bibitem [{\citenamefont {McCarthy}\ \emph {et~al.}(2024)\citenamefont
  {McCarthy}, \citenamefont {Pirvu}, \citenamefont {Hill}, \citenamefont
  {Huang}, \citenamefont {Johnson},\ and\ \citenamefont
  {Rogers}}]{McCarthy:2024ozh}%
  \BibitemOpen
  \bibfield  {author} {\bibinfo {author} {\bibfnamefont {F.}~\bibnamefont
  {McCarthy}}, \bibinfo {author} {\bibfnamefont {D.}~\bibnamefont {Pirvu}},
  \bibinfo {author} {\bibfnamefont {J.~C.}\ \bibnamefont {Hill}}, \bibinfo
  {author} {\bibfnamefont {J.}~\bibnamefont {Huang}}, \bibinfo {author}
  {\bibfnamefont {M.~C.}\ \bibnamefont {Johnson}}, \ and\ \bibinfo {author}
  {\bibfnamefont {K.~K.}\ \bibnamefont {Rogers}},\ }\href {\doibase
  10.1103/PhysRevLett.133.141003} {\bibfield  {journal} {\bibinfo  {journal}
  {Phys. Rev. Lett.}\ }\textbf {\bibinfo {volume} {133}},\ \bibinfo {pages}
  {141003} (\bibinfo {year} {2024})},\ \Eprint
  {http://arxiv.org/abs/2406.02546} {arXiv:2406.02546 [hep-ph]} \BibitemShut
  {NoStop}%
\bibitem [{\citenamefont {P{\textasciicircum}{\i}rvu}\ \emph
  {et~al.}(2024)\citenamefont {P{\textasciicircum}{\i}rvu}, \citenamefont
  {Huang},\ and\ \citenamefont {Johnson}}]{Pirvu:2023lch}%
  \BibitemOpen
  \bibfield  {author} {\bibinfo {author} {\bibfnamefont {D.}~\bibnamefont
  {P{\textasciicircum}{\i}rvu}}, \bibinfo {author} {\bibfnamefont
  {J.}~\bibnamefont {Huang}}, \ and\ \bibinfo {author} {\bibfnamefont {M.~C.}\
  \bibnamefont {Johnson}},\ }\href {\doibase 10.1088/1475-7516/2024/01/019}
  {\bibfield  {journal} {\bibinfo  {journal} {JCAP}\ }\textbf {\bibinfo
  {volume} {01}},\ \bibinfo {pages} {019} (\bibinfo {year} {2024})},\ \Eprint
  {http://arxiv.org/abs/2307.15124} {arXiv:2307.15124 [hep-ph]} \BibitemShut
  {NoStop}%
\bibitem [{\citenamefont {Aramburo-Garcia}\ \emph {et~al.}(2024)\citenamefont
  {Aramburo-Garcia}, \citenamefont {Bondarenko}, \citenamefont {Boyarsky},
  \citenamefont {Kashko}, \citenamefont {Pradler}, \citenamefont {Sokolenko},
  \citenamefont {Kugel}, \citenamefont {Schaller},\ and\ \citenamefont
  {Schaye}}]{Aramburo-Garcia:2024cbz}%
  \BibitemOpen
  \bibfield  {author} {\bibinfo {author} {\bibfnamefont {A.}~\bibnamefont
  {Aramburo-Garcia}}, \bibinfo {author} {\bibfnamefont {K.}~\bibnamefont
  {Bondarenko}}, \bibinfo {author} {\bibfnamefont {A.}~\bibnamefont
  {Boyarsky}}, \bibinfo {author} {\bibfnamefont {P.}~\bibnamefont {Kashko}},
  \bibinfo {author} {\bibfnamefont {J.}~\bibnamefont {Pradler}}, \bibinfo
  {author} {\bibfnamefont {A.}~\bibnamefont {Sokolenko}}, \bibinfo {author}
  {\bibfnamefont {R.}~\bibnamefont {Kugel}}, \bibinfo {author} {\bibfnamefont
  {M.}~\bibnamefont {Schaller}}, \ and\ \bibinfo {author} {\bibfnamefont
  {J.}~\bibnamefont {Schaye}},\ }\href {\doibase 10.1088/1475-7516/2024/11/049}
  {\bibfield  {journal} {\bibinfo  {journal} {JCAP}\ }\textbf {\bibinfo
  {volume} {11}},\ \bibinfo {pages} {049} (\bibinfo {year} {2024})},\ \Eprint
  {http://arxiv.org/abs/2405.05104} {arXiv:2405.05104 [astro-ph.CO]}
  \BibitemShut {NoStop}%
\bibitem [{\citenamefont {Mirasola}\ \emph {et~al.}(2025)\citenamefont
  {Mirasola} \emph {et~al.}}]{Mirasola:2025car}%
  \BibitemOpen
  \bibfield  {author} {\bibinfo {author} {\bibfnamefont {L.}~\bibnamefont
  {Mirasola}} \emph {et~al.},\ }\href {\doibase 10.1103/PhysRevD.111.084032}
  {\bibfield  {journal} {\bibinfo  {journal} {Phys. Rev. D}\ }\textbf {\bibinfo
  {volume} {111}},\ \bibinfo {pages} {084032} (\bibinfo {year} {2025})},\
  \Eprint {http://arxiv.org/abs/2501.02052} {arXiv:2501.02052 [gr-qc]}
  \BibitemShut {NoStop}%
\bibitem [{\citenamefont {Holdom}(1986)}]{Holdom1986}%
  \BibitemOpen
  \bibfield  {author} {\bibinfo {author} {\bibfnamefont {B.}~\bibnamefont
  {Holdom}},\ }\href {\doibase 10.1016/0370-2693(86)91377-8} {\bibfield
  {journal} {\bibinfo  {journal} {Physics Letters B}\ }\textbf {\bibinfo
  {volume} {166}},\ \bibinfo {pages} {196} (\bibinfo {year}
  {1986})}\BibitemShut {NoStop}%
\bibitem [{\citenamefont {Caputo}\ \emph {et~al.}(2025)\citenamefont {Caputo},
  \citenamefont {Franciolini},\ and\ \citenamefont {Witte}}]{Caputo:2025oap}%
  \BibitemOpen
  \bibfield  {author} {\bibinfo {author} {\bibfnamefont {A.}~\bibnamefont
  {Caputo}}, \bibinfo {author} {\bibfnamefont {G.}~\bibnamefont {Franciolini}},
  \ and\ \bibinfo {author} {\bibfnamefont {S.~J.}\ \bibnamefont {Witte}},\
  }\href@noop {} {\  (\bibinfo {year} {2025})},\ \Eprint
  {http://arxiv.org/abs/2507.21788} {arXiv:2507.21788 [hep-ph]} \BibitemShut
  {NoStop}%
\bibitem [{\citenamefont {Khalaf}\ \emph {et~al.}(2024)\citenamefont {Khalaf},
  \citenamefont {Kuflik}, \citenamefont {Lenoci},\ and\ \citenamefont
  {Stone}}]{Khalaf:2024nwc}%
  \BibitemOpen
  \bibfield  {author} {\bibinfo {author} {\bibfnamefont {M.}~\bibnamefont
  {Khalaf}}, \bibinfo {author} {\bibfnamefont {E.}~\bibnamefont {Kuflik}},
  \bibinfo {author} {\bibfnamefont {A.}~\bibnamefont {Lenoci}}, \ and\ \bibinfo
  {author} {\bibfnamefont {N.~C.}\ \bibnamefont {Stone}},\ }\href@noop {} {\
  (\bibinfo {year} {2024})},\ \Eprint {http://arxiv.org/abs/2408.16051}
  {arXiv:2408.16051 [astro-ph.CO]} \BibitemShut {NoStop}%
\bibitem [{\citenamefont {Mummery}\ \emph {et~al.}(2025)\citenamefont
  {Mummery}, \citenamefont {Jiang}, \citenamefont {Ingram}, \citenamefont
  {Fabian},\ and\ \citenamefont {Rule}}]{Mummery:2025tvq}%
  \BibitemOpen
  \bibfield  {author} {\bibinfo {author} {\bibfnamefont {A.}~\bibnamefont
  {Mummery}}, \bibinfo {author} {\bibfnamefont {J.}~\bibnamefont {Jiang}},
  \bibinfo {author} {\bibfnamefont {A.}~\bibnamefont {Ingram}}, \bibinfo
  {author} {\bibfnamefont {A.}~\bibnamefont {Fabian}}, \ and\ \bibinfo {author}
  {\bibfnamefont {J.}~\bibnamefont {Rule}},\ }\href@noop {} {\  (\bibinfo
  {year} {2025})},\ \Eprint {http://arxiv.org/abs/2505.13119} {arXiv:2505.13119
  [astro-ph.HE]} \BibitemShut {NoStop}%
\bibitem [{\citenamefont {Sun}\ \emph {et~al.}(2020{\natexlab{b}})\citenamefont
  {Sun}, \citenamefont {Brito},\ and\ \citenamefont {Isi}}]{Sun:2019mqb}%
  \BibitemOpen
  \bibfield  {author} {\bibinfo {author} {\bibfnamefont {L.}~\bibnamefont
  {Sun}}, \bibinfo {author} {\bibfnamefont {R.}~\bibnamefont {Brito}}, \ and\
  \bibinfo {author} {\bibfnamefont {M.}~\bibnamefont {Isi}},\ }\href {\doibase
  10.1103/PhysRevD.101.063020} {\bibfield  {journal} {\bibinfo  {journal}
  {Phys. Rev. D}\ }\textbf {\bibinfo {volume} {101}},\ \bibinfo {pages}
  {063020} (\bibinfo {year} {2020}{\natexlab{b}})},\ \bibinfo {note} {[Erratum:
  Phys.Rev.D 102, 089902 (2020)]},\ \Eprint {http://arxiv.org/abs/1909.11267}
  {arXiv:1909.11267 [gr-qc]} \BibitemShut {NoStop}%
\bibitem [{\citenamefont {Abac}\ \emph
  {et~al.}(2025{\natexlab{c}})\citenamefont {Abac} \emph
  {et~al.}}]{LIGOScientific:2025csr}%
  \BibitemOpen
  \bibfield  {author} {\bibinfo {author} {\bibfnamefont {A.~G.}\ \bibnamefont
  {Abac}} \emph {et~al.} (\bibinfo {collaboration} {LIGO Scientific, VIRGO,
  KAGRA}),\ }\href@noop {} {\  (\bibinfo {year} {2025}{\natexlab{c}})},\
  \Eprint {http://arxiv.org/abs/2509.07352} {arXiv:2509.07352 [gr-qc]}
  \BibitemShut {NoStop}%
\bibitem [{\citenamefont {Yoshino}\ and\ \citenamefont
  {Kodama}(2015)}]{Yoshino:2015nsa}%
  \BibitemOpen
  \bibfield  {author} {\bibinfo {author} {\bibfnamefont {H.}~\bibnamefont
  {Yoshino}}\ and\ \bibinfo {author} {\bibfnamefont {H.}~\bibnamefont
  {Kodama}},\ }\href {\doibase 10.1088/0264-9381/32/21/214001} {\bibfield
  {journal} {\bibinfo  {journal} {Class. Quant. Grav.}\ }\textbf {\bibinfo
  {volume} {32}},\ \bibinfo {pages} {214001} (\bibinfo {year} {2015})},\
  \Eprint {http://arxiv.org/abs/1505.00714} {arXiv:1505.00714 [gr-qc]}
  \BibitemShut {NoStop}%
\bibitem [{\citenamefont {Siemonsen}\ and\ \citenamefont
  {East}(2020)}]{Siemonsen:2019ebd}%
  \BibitemOpen
  \bibfield  {author} {\bibinfo {author} {\bibfnamefont {N.}~\bibnamefont
  {Siemonsen}}\ and\ \bibinfo {author} {\bibfnamefont {W.~E.}\ \bibnamefont
  {East}},\ }\href {\doibase 10.1103/PhysRevD.101.024019} {\bibfield  {journal}
  {\bibinfo  {journal} {Phys. Rev. D}\ }\textbf {\bibinfo {volume} {101}},\
  \bibinfo {pages} {024019} (\bibinfo {year} {2020})},\ \Eprint
  {http://arxiv.org/abs/1910.09476} {arXiv:1910.09476 [gr-qc]} \BibitemShut
  {NoStop}%
\bibitem [{\citenamefont {Dolan}(2018)}]{Dolan:2018dqv}%
  \BibitemOpen
  \bibfield  {author} {\bibinfo {author} {\bibfnamefont {S.~R.}\ \bibnamefont
  {Dolan}},\ }\href {\doibase 10.1103/PhysRevD.98.104006} {\bibfield  {journal}
  {\bibinfo  {journal} {Phys. Rev. D}\ }\textbf {\bibinfo {volume} {98}},\
  \bibinfo {pages} {104006} (\bibinfo {year} {2018})},\ \Eprint
  {http://arxiv.org/abs/1806.01604} {arXiv:1806.01604 [gr-qc]} \BibitemShut
  {NoStop}%
\bibitem [{\citenamefont {Baumann}\ \emph {et~al.}(2019)\citenamefont
  {Baumann}, \citenamefont {Chia}, \citenamefont {Stout},\ and\ \citenamefont
  {ter Haar}}]{Baumann:2019eav}%
  \BibitemOpen
  \bibfield  {author} {\bibinfo {author} {\bibfnamefont {D.}~\bibnamefont
  {Baumann}}, \bibinfo {author} {\bibfnamefont {H.~S.}\ \bibnamefont {Chia}},
  \bibinfo {author} {\bibfnamefont {J.}~\bibnamefont {Stout}}, \ and\ \bibinfo
  {author} {\bibfnamefont {L.}~\bibnamefont {ter Haar}},\ }\href {\doibase
  10.1088/1475-7516/2019/12/006} {\bibfield  {journal} {\bibinfo  {journal}
  {JCAP}\ }\textbf {\bibinfo {volume} {12}},\ \bibinfo {pages} {006} (\bibinfo
  {year} {2019})},\ \Eprint {http://arxiv.org/abs/1908.10370} {arXiv:1908.10370
  [gr-qc]} \BibitemShut {NoStop}%
\bibitem [{\citenamefont {Combi}\ \emph {et~al.}()\citenamefont {Combi},
  \citenamefont {Wong}, \citenamefont {Siemonsen},\ and\ \citenamefont
  {Huang}}]{inprep}%
  \BibitemOpen
  \bibfield  {author} {\bibinfo {author} {\bibfnamefont {L.}~\bibnamefont
  {Combi}}, \bibinfo {author} {\bibfnamefont {G.}~\bibnamefont {Wong}},
  \bibinfo {author} {\bibfnamefont {N.}~\bibnamefont {Siemonsen}}, \ and\
  \bibinfo {author} {\bibfnamefont {J.}~\bibnamefont {Huang}},\ }\href@noop {}
  {\bibinfo  {journal} {in prep}\ }\BibitemShut {NoStop}%
\bibitem [{\citenamefont {Omiya}\ \emph {et~al.}(2024)\citenamefont {Omiya},
  \citenamefont {Takahashi}, \citenamefont {Tanaka},\ and\ \citenamefont
  {Yoshino}}]{Omiya:2024xlz}%
  \BibitemOpen
\bibfield  {journal} {  }\bibfield  {author} {\bibinfo {author} {\bibfnamefont
  {H.}~\bibnamefont {Omiya}}, \bibinfo {author} {\bibfnamefont
  {T.}~\bibnamefont {Takahashi}}, \bibinfo {author} {\bibfnamefont
  {T.}~\bibnamefont {Tanaka}}, \ and\ \bibinfo {author} {\bibfnamefont
  {H.}~\bibnamefont {Yoshino}},\ }\href {\doibase 10.1103/PhysRevD.110.044002}
  {\bibfield  {journal} {\bibinfo  {journal} {Phys. Rev. D}\ }\textbf {\bibinfo
  {volume} {110}},\ \bibinfo {pages} {044002} (\bibinfo {year} {2024})},\
  \Eprint {http://arxiv.org/abs/2404.16265} {arXiv:2404.16265 [gr-qc]}
  \BibitemShut {NoStop}%
\bibitem [{\citenamefont {Abbott}\ \emph {et~al.}(2021)\citenamefont {Abbott}
  \emph {et~al.}}]{LIGOScientific:2020ibl}%
  \BibitemOpen
  \bibfield  {author} {\bibinfo {author} {\bibfnamefont {R.}~\bibnamefont
  {Abbott}} \emph {et~al.} (\bibinfo {collaboration} {LIGO Scientific,
  Virgo}),\ }\href {\doibase 10.1103/PhysRevX.11.021053} {\bibfield  {journal}
  {\bibinfo  {journal} {Phys. Rev. X}\ }\textbf {\bibinfo {volume} {11}},\
  \bibinfo {pages} {021053} (\bibinfo {year} {2021})},\ \Eprint
  {http://arxiv.org/abs/2010.14527} {arXiv:2010.14527 [gr-qc]} \BibitemShut
  {NoStop}%
\bibitem [{\citenamefont {Varma}\ \emph {et~al.}(2019)\citenamefont {Varma},
  \citenamefont {Field}, \citenamefont {Scheel}, \citenamefont {Blackman},
  \citenamefont {Gerosa}, \citenamefont {Stein}, \citenamefont {Kidder},\ and\
  \citenamefont {Pfeiffer}}]{Varma:2019csw}%
  \BibitemOpen
  \bibfield  {author} {\bibinfo {author} {\bibfnamefont {V.}~\bibnamefont
  {Varma}}, \bibinfo {author} {\bibfnamefont {S.~E.}\ \bibnamefont {Field}},
  \bibinfo {author} {\bibfnamefont {M.~A.}\ \bibnamefont {Scheel}}, \bibinfo
  {author} {\bibfnamefont {J.}~\bibnamefont {Blackman}}, \bibinfo {author}
  {\bibfnamefont {D.}~\bibnamefont {Gerosa}}, \bibinfo {author} {\bibfnamefont
  {L.~C.}\ \bibnamefont {Stein}}, \bibinfo {author} {\bibfnamefont {L.~E.}\
  \bibnamefont {Kidder}}, \ and\ \bibinfo {author} {\bibfnamefont {H.~P.}\
  \bibnamefont {Pfeiffer}},\ }\href {\doibase 10.1103/PhysRevResearch.1.033015}
  {\bibfield  {journal} {\bibinfo  {journal} {Phys. Rev. Research.}\ }\textbf
  {\bibinfo {volume} {1}},\ \bibinfo {pages} {033015} (\bibinfo {year}
  {2019})},\ \Eprint {http://arxiv.org/abs/1905.09300} {arXiv:1905.09300
  [gr-qc]} \BibitemShut {NoStop}%
\end{thebibliography}%
\appendix
\section{Comparison to existing constraints}
\label{app:work_comp}
Our constraints are complementary to existing electromagnetic and GW searches for ultralight bosons. By directly leveraging the measured component spins of individual binary BH merger systems, we place limits on superradiant spin-down that are independent of assumptions about the BH population or accretion physics. 
Consequently, directly comparing all constraints requires carefully accounting for the differing underlying assumptions. Here, we provide a brief qualitative summary of the main existing electromagnetic and GW constraints to place our results in context.

Using spin measurements from continuum-fitting (thermal disk) and Fe–K$\alpha$ reflection (relativistically broadened iron lines), several works derived Regge-plane exclusions. Reference~\cite{Arvanitaki:2014wva} showed that observed near-maximal spins in several stellar and supermassive BHs exclude scalar masses $\sim 6\times10^{-13}$–$2\times10^{-11}\,\mathrm{eV}$, with the strongest (and least model-dependent) constraints coming from stellar-mass BHs. Reference~\cite{Stott:2018opm} extended these exclusions to multifield, axionlike particle (ALP) scenarios. 
More recently, Refs.~\cite{Baryakhtar:2020gao,Hoof:2024quk} extended the mass range for axions excluding $\sim 3\times10^{-13}$–$6\times10^{-12}\,\mathrm{eV}$ using several stellar mass X-ray binaries.
For vectors and higher azimuthal modes, Ref.~\cite{Baryakhtar:2017ngi} derived constraints using BH X-ray spin measurements that disfavor vector masses approximately in the range $\sim 5\times10^{-14}$–$2\times10^{-11}\,\mathrm{eV}$ for stellar BHs (and, with larger astrophysical uncertainties, in the range $\sim 6\times10^{-20}$–$2\times10^{-17}\,\mathrm{eV}$ using supermassive BHs). Reference~\cite{Cardoso:2018tly} and follow-up studies obtained comparable limits for dark photons with masses $\sim 10^{-13}$–$3\times10^{-12}\,\mathrm{eV}$ and for higher-$m$ ALP modes in the mass range of $\sim 6\times10^{-13}$–$10^{-11}\,\mathrm{eV}$, noting that supermassive BH spin distribution studies probe even lighter bosons at $\sim10^{-19}\,$eV but with larger model dependence. More recently, Khalaf et al.~\cite{Khalaf:2024nwc} proposed directly comparing continuum-fit and reflection spins of the same source as a robust observational test for boson clouds.
However, these electromagnetic-based constraints carry substantial modeling uncertainties, primarily due to assumptions about the accretion physics and spin-inference systematics. (See, e.g., Ref.~\cite{Mummery:2025tvq}).

GW observations offer complementary probes that are not sensitive to uncertainties in accretion modeling. Population-level analyses, such as the hierarchical Bayesian studies of Refs.~\cite{Ng:2019jsx,Ng:2020ruv}, combine spin measurements from multiple binary BH mergers to search for evidence of superradiant spin-down. Those analyses assume hierarchical priors on the underlying spin and mass distributions and that the component spins evolve only through the superradiant instability of noninteracting scalar fields, neglecting accretion or environmental effects. Applying this framework to the GWTC-2 catalog disfavors scalar boson masses in the range 
$[1.3, 2.7]\times 10^{-13}$~eV, driven by the high-spin events GW190517 and (to a lesser extent) GW190412. Our analysis also includes GW190517 and finds comparable constraints for scalars under the same age assumptions, but takes a different approach by constraining the instability directly from the measured component spins, independent of population model or priors.
Reference~\cite{Ghosh:2021zuf} also uses the one-sigma mass-spin posteriors for the primary in GW190517 to exclude vectors in the range $1.7\times 10^{-14}$-$7.6 \times10^{-13}$ eV, assuming $T_{\rm age}=10^7$ yr. (That reference also analyzes GW190426\_152155, which we do not consider as it has a quoted false alarm rate of $1.4$ yr${}^{-1}$~\cite{GWTC-2.1}.)
Our analysis finds a very similar lower bound for this event under the same assumption, though a slightly higher upper bound for the mass exclusion, which can be attributed to the inclusion of higher azimuthal modes. 
Searches for gravitational radiation from boson clouds as stochastic backgrounds have yielded constraints on scalars in the range of
$[2.0, 3.8]\times 10^{-13}$~eV ~\cite{Tsukada:2018mbp}, and vectors in the range of 
$[0.8, 6.0]\times 10^{-13}$~eV ~\cite{Tsukada:2020lgt}. These constraints rely on assumptions about the underlying BH population, and shrink or disappear if one assumes BHs are predominantly low spin.

Blind all-sky continuous-wave searches translate strain upper limits into exclusions in the BH mass–boson mass plane under specific assumptions about the distributions of BH mass, age, distance, and spin. The limits obtained from all-sky continuous-wave searches in earlier observing runs disfavor scalar masses in the range $\sim10^{-13}$–$10^{-11}\,$eV ~\cite{Palomba2019, Dergachev2019}, while population-based analyses of isolated Galactic BHs exclude $\sim2\times10^{-13}$--$2.5\times10^{-12}$~\cite{Zhu2020}.
The all-sky search tailored to scalar boson signals in the third LIGO-Virgo-KAGRA observing run produced strain upper limits $\sim10^{-25}$ at $\sim130\,$Hz, yielding exclusion contours once source assumptions are specified~\cite{O3_all-sky_scalar_bosons}. 
A directed search in Advanced LIGO's second observing run targeting the BH in Cygnus~X-1, yields constraints under the assumptions of a near-extremal birth spin and an age of $\sim 10^{5}$~yr. 
Assuming purely gravitational interactions, scalar masses in the range $[6.3, 13.2]\times 10^{-13}$~eV are disfavored~\cite{O2_CygnusX1_scalar_bosons,Sun:2019mqb}. 
With non-negligible self-interaction strengths $\lesssim 10^{18}$ GeV, no mass range can be confidently excluded \cite{Collaviti2024}. 
A recent study using data from the first part of O4 excluded a vector boson mass range of $[0.85, 1.59]\times 10^{-13}$~eV for Cygnus X-1, assuming an age of $\sim 6 \times 10^{6}$ yr~\cite{LIGOScientific:2025csr}. 
Vector searches targeting the merger remnant black holes of GW231123 and GW230814 in O4 yielded marginally (30\% confidence) disfavored vector mass ranges of $[0.94, 1.08] \times 10^{-13}$~eV and $[2.75, 3.28] \times 10^{-13}$~eV, respectively~\cite{LIGOScientific:2025csr}.

\section{Calculating superradiant spin-down}
\label{app:spindown_calc}
In this section, we give more details on the calculation of how a BH would be spun down due to the (purely gravitational) superradiance instability. As a starting point, we use the boson cloud mass $e$-folding time $\tau$ and real angular frequency $\omega_R$ calculated in the test-field limit on a Kerr BH background, as described in Ref.~\cite{Siemonsen:2022yyf}. We have extended the relativistic calculations presented there to include azimuthal modes with $m=3$, 4, and 5 for the vector case, as described in Appendix~\ref{app:higher_m}. Thus, for $m\leq 2$ in the scalar case, and $m\leq 5$ in the vector case, we use the relativistic instability results. For larger $m$ values, we use the nonrelativistic approximations, though these do not contribute significantly to the constraints.

When considering the spin-down of a BH with initial mass $M_i$ and angular momentum $J_i$ over a timescale $T_{\rm age}$, we begin by finding the azimuthal number of the fastest growing superradiantly unstable mode.\footnote{We focus exclusively on the zeroth radial overtone, which is not necessarily the most unstable mode in extreme regions of the parameter space (see, e.g., Fig. 3 in both Ref.~\cite{Yoshino:2015nsa} and Ref.~\cite{Siemonsen:2019ebd}). 
This underestimates the growth rate and hence is a conservative assumption.} Typically, this is given by the smallest $m$ value for which the superradiant condition $\omega_R<m \Omega_{\rm BH}$ is satisfied, where $\Omega_{\rm BH}$ is the horizon frequency of the BH. We determine the mass of the boson cloud $M_c$ at the saturation of the superradiant instability by solving\footnote{Here we use $G=c=1$ units.}
\begin{equation}
\omega_R(M_f,J_f)= m \Omega_{\rm BH}(M_f,J_f),
\end{equation}
with the final BH mass and angular momentum 
\begin{equation}\label{eqn:m_f}
M_f=M_i-M_c \text{ and } J_f=J_i-\frac{m}{\omega_R}M_c \ .
\end{equation}
We include the change in the BH mass due to the instability, though in general
this is a small correction.
We assume that the timescale for the cloud to grow to saturation is given by $T=\tau \log(M_c/m_b)$, i.e., the cloud grows from a single boson. If $T<T_{\rm age}$, we then repeat the spin-down calculation with $M_f$, $J_f$, and using $T_{\rm age}-T$ as the new spin-down timescale to check if higher azimuthal modes can spin down the BH further. Otherwise, we assume that the cloud
mass has grown exponentially to the corresponding fraction of its saturation value, given by $M_c\exp[(T_{\rm age}-T)/\tau]$, and calculate the final BH mass $M_f$ and dimensionless spin $\chi_f$ as in Eq.~\eqref{eqn:m_f}.

\begin{figure*}[htbp]
    \centering
     \includegraphics[width=0.8\linewidth]{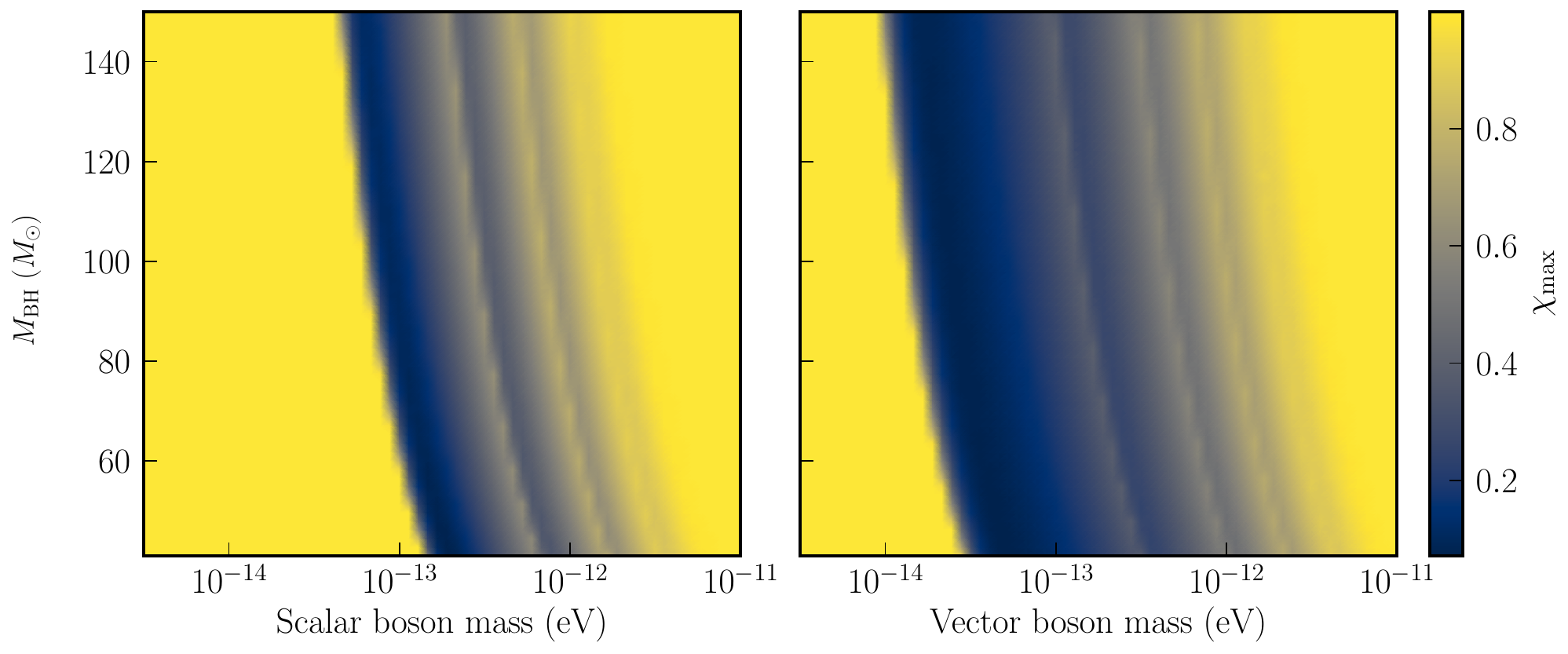}
    \caption{
    Maximum dimensionless spin allowed as a function of boson mass and final BH mass when assuming a BH age of $T_{\rm age}=10^5$ years in the presence of a scalar (left) and vector (right).
    }
    \label{fig:chi_max}
\end{figure*}

By applying the above calculation to an ensemble of BHs, one can then determine the maximum value of the dimensionless spin $\chi_{\rm max}(M_f, m_b, T_{\rm age})$ for given values of $m_b$ and $T_{\rm age}$, such that a BH of any initial mass and spin values would, after the superradiant instability, end up with a final BH spin $\chi_f\leq \chi_{\rm max}(M_f, m_b, T_{\rm age})$. In Fig.~\ref{fig:chi_max}, we plot $\chi_{\rm max}(M_f,m_{b},T_{\rm age}=10^5 {\rm years})$ for scalar and vectors. In the figure, the shorter instability timescale of the vector boson instability is apparent from the smaller values of $\chi_{\rm max}$ compared to the scalar case. One also notices a characteristic ``striping'' moving from lower to higher boson masses that is due to the contributions from the higher azimuthal number unstable modes.

\section{Higher azimuthal vector modes}
\label{app:higher_m}

Here we provide further details on the frequencies $\omega_R$ and growth rates $\omega_I=\tau^{-1}/2$ used for the higher-order azimuthal modes $m=3$, 4, and $5$ of vector superradiance. These are obtained by solving the massive vector field equation on a Kerr background spacetime for a given choice of spin $\chi$ and gravitational fine structure constant $\alpha=M m_b$; in this appendix, we set $G=c=\hbar=1$. Imposing ingoing boundary conditions at the horizon and vanishing field at spatial infinity, the vector field equation can be reduced to a complex eigenvalue problem involving practically only a single radial ordinary differential equation. Both numerical and analytic solution techniques have been employed to obtain the eigenvalue $\omega=\omega_R+i\omega_I$. While analytic techniques are applicable in the $\alpha\ll 1$ regime, numerical methods are most accurate for $\alpha\sim \mathcal{O}(1)$, rendering the two highly complementary. In order to obtain the most accurate estimates for $\omega$, we utilize the same procedure as previously used for constructing \texttt{SuperRad}~\cite{Siemonsen:2022yyf}. In particular, we use the methods of Refs.~\cite{Siemonsen:2019ebd,Dolan:2018dqv} to numerically solve the full massive vector field equations on the Kerr background on a grid covering the regions $\alpha_0<\alpha<\alpha_{\rm sat.}$, where $\alpha_0=0.35$, 0.5, and 0.6 for the $m=3$, 4, and 5 modes, respectively,
and $\alpha_{\rm sat.}$ is the corresponding saturation value of the fundamental mode\footnote{Recall, we ignore effects originating from higher radial overtones.} for each value of $m$. We cover
dimensionless spin range $\chi\in[0.6,0.995]$. Within this region of the parameter space, the growth rate is then obtained by simple interpolation, while outside this region, the rates are obtained from a fit to these numerical data, which recovers the known nonrelativistic analytic estimates for $\alpha\ll 1$ \cite{Baryakhtar:2017ngi,Baumann:2019eav}, as described in detail in Ref.~\cite{Siemonsen:2022yyf}. 

\begin{figure}[t]
\includegraphics[width=1\linewidth]{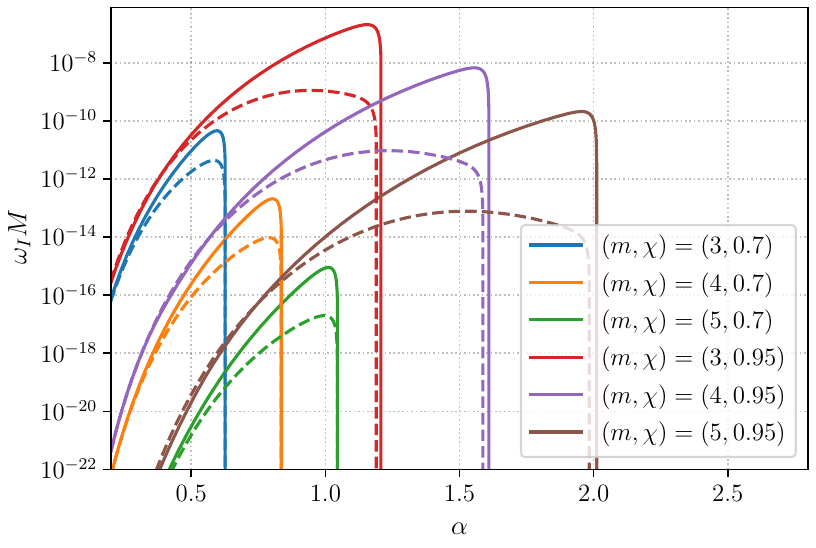}
\caption{The growth rate $\omega_I$ of the higher order modes for two selected BH spins $\chi$ as a function of $\alpha$. Solid curves show the relativistically correct estimates, while the dashed curves show the nonrelativistic analytic results, expected to be accurate only in the $\alpha\ll 1$ regime.}
\label{fig:modes}
\end{figure}

As a point of comparison, and to highlight the impact of relativistic effects, in Fig.~\ref{fig:modes} we show both the purely analytic nonrelativistic estimates together with those obtained through the procedure described above. By construction, the two agree in the $\alpha\ll 1$ regime, while both for large spins and large $\alpha$, the nonrelativistic estimates underestimate the true growth rates by orders of magnitude. For $0.95<\chi<1$, the difference is even greater than that shown in Fig.~\ref{fig:modes} for $\alpha\sim \mathcal{O}(1)$.
\begin{figure}[t]
\includegraphics[width=1\linewidth]{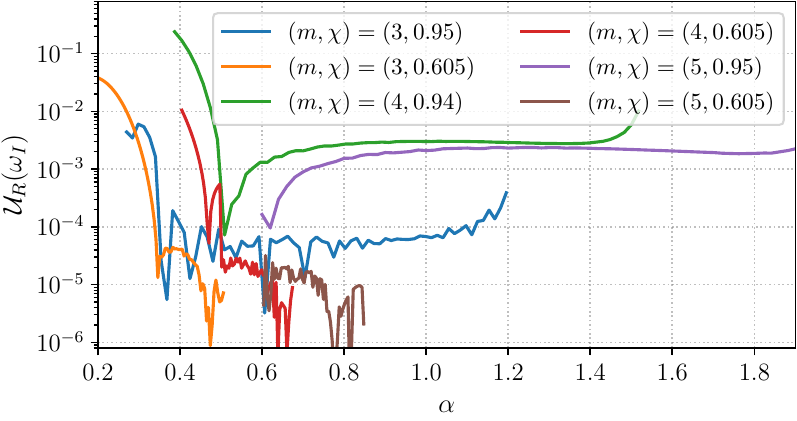}
\caption{The relative errors, $\mathcal{U}_R$, defined in the text, of the growth rates $\omega_I$ as a function of $\alpha$ for two selected spins $\chi$.}
\label{fig:mode_acc}
\end{figure}
Finally, we estimate the error of our methods by numerically computing the growth rates at a moderate and a large BH spin value different from those used in the above procedure. This allows us to both obtain an estimate for the interpolation error (likely dominating the total error budget in the relativistic regime) for $\alpha_0<\alpha$ and a rough measure of the accuracy of the fit for $\alpha<\alpha_0$. We then define $\mathcal{U}_R(x)=|x-x_{\rm numerical}|/x$ to determine the relative error between the numerically computed rates and those obtained from the above interpolation/fitting procedure. As shown in Fig.~\ref{fig:mode_acc}, the interpolation error in the region $\alpha_0<\alpha$ is below the 10\% level for all tested spins and modes. Since numerical data can be reliably computed only for small $m$ and large $\alpha$, we were able to obtain estimates of the accuracy of the fit for $\omega_I$, i.e., $\mathcal{U}_R(\omega_I)$, only for $m=3$ and $m=4$. This accuracy for $\alpha<\alpha_0$ is at, or worse than, the 10\% level. This is comparable to, or worse than, the equivalent measures for the $m=1$ and $2$ modes shown in Fig. 14 of Ref.~\cite{Siemonsen:2022yyf}; recall from above, for $\alpha\rightarrow 0$, this relative difference will decrease to zero. Reduced overlapping regions of validity for the numerical methods, on the one hand, and nonrelativistic analytic estimates, on the other hand, lead to less accurate fits for $\alpha\lesssim \alpha_0$ and higher azimuthal modes, which is reflected in measures such as the one shown in Fig.~\ref{fig:mode_acc}.

\section{Boson models with interactions}
\label{app:interact}
In this section, we provide additional details on how constraints
are obtained in boson models with additional (nongravitational) interactions. We again make use of $G=c=\hbar=1$ units.

\subsection{Kinetic mixing}
In the dark photon model, a nonvanishing kinetic mixing strength $\varepsilon$ may impact the superradiance process due to an additional dissipation channel through electromagnetic emission \cite{Siemonsen:2022ivj}. In particular, if the electromagnetic energy flux towards large distances matches the energy flux across the horizon, then the exponential growth is halted and the BH spin-down is stopped. This dissipation channel is powered by a pair plasma produced through a photon-assisted pair production cascade. In order to map the constraints obtained above on the vector boson's mass into an upper bound on the kinetic mixing strength $\varepsilon$, we use the results of Ref.~\cite{Siemonsen:2022ivj} and construct mappings in direct analogy to the approach taken in Ref.~\cite{Jones:2024fpg}. 

The relevant timescales associated with dark photon superradiance are (i) the superradiance instability e-folding time $\tau$, (ii) the pair production rate $\Gamma_\pm$, (iii) the cloud's light crossing timescale $\tau_{\rm LC}$, and (iv) the electromagnetic dissipation timescale $\tau_{\rm EM}$. The pair plasma is produced efficiently only if\footnote{The total pair production time is parametrically shorter than the instability $e$-folding time for the $m=1$ \cite{Siemonsen:2022ivj}.} $\Gamma_\pm>\tau_{\rm LC}^{-1}=\alpha m_b$, whereas the resulting electromagnetic dissipation quenches the superradiance instability only if\footnote{We conservatively ignore any linear-in-time BH spin-down, when $\tau>\tau_{\rm EM}$ is satisfied.} $\tau>\tau_{\rm EM}$. The requirement $\Gamma_\pm>\tau_{\rm LC}^{-1}$ implies that a pair plasma is produced only for
\begin{align}
\varepsilon\gtrsim\varepsilon_\pm= \frac{\sqrt{2 m_b}m_e^{3/2}}{e|\textbf{E}'_{\rm SR}|},
\label{eq:pairprodcrit}
\end{align}
where we drop logarithmic corrections and introduce the electron mass $m_e$, and charge $e$, as well as the cloud's electric field $\textbf{E}'_{\rm SR}$. The maximal electric field strength of the dark photon cloud depends exponentially on the azimuthal mode number:\footnote{Notice, the location of the maximum of the field strength shifts as $r_{\rm max}\sim m M/\alpha^2$ away from the central BH for $m>1$, likely suppressing relativistic effects.}
\begin{align}
\max (\textbf{E}'_{\rm SR} M)^2=C_m \alpha^6\frac{M_c}{M},
\end{align}
where
\begin{align}
\begin{aligned}
C_1=\pi^{-1}, \quad C_2=(16 e^2\pi)^{-1}, \quad C_3 = 4(81e^4\pi)^{-1},\\
C_4 = 81 (1024 e^6\pi)^{-1}, \quad C_5 = 1024 (5625 e^8\pi)^{-1}.
\end{aligned}
\end{align}
We derive these coefficients from the solutions found in Ref.~\cite{Baryakhtar:2017ngi}. For $m=1$, there is a relativistic (albeit gauge-dependent) enhancement of the electric field strength for $\alpha\sim \mathcal{O}(1)$ of $\sim 10$ \cite{inprep}. While this relativistic enhancement is likely suppressed for $m>1$, we conservatively include a factor of $10$ in $\varepsilon_\pm$ for all azimuthal modes. Now, if Eq.~\eqref{eq:pairprodcrit} is satisfied for a given azimuthal mode $m$ and for $M_c\lesssim M_c^{\rm sat}$ (where $M_c^{\rm sat}$ is the cloud's mass at gravitational saturation), then a pair plasma is formed, which has the potential to disrupt the superradiant growth if $\tau>\tau_{\rm EM}$ also holds. We define $\tau=M_c/\dot{E}_{\rm BH}$ (as above) and $\tau_{\rm EM}=M_c/L_{\rm EM}$, where the electromagnetic power output\footnote{Notice also, while this expression was obtained for a superradiance saturated cloud state in Ref.~\cite{Siemonsen:2022ivj}, we expect the unsaturated field configuration to exhibit similar electromagnetic power outputs.} $L_{\rm EM}=\varepsilon^2F(\alpha)(M_c/M)$ with $F(\alpha)=0.131\alpha-0.188\alpha^2$ is known only for the $m=1$ mode, and $\dot{E}_{\rm BH}$ is the energy flux of the unstable field configuration across the horizon. This implies that $\tau >\tau_{\rm EM}$ if
\begin{align}
\varepsilon\gtrsim\varepsilon_{\rm diss}:=\left(\frac{M}{F(\alpha)\tau}\right)^{1/2}.
\end{align}
Hence, the BH spin measurements apply to all kinetic mixing values with
\begin{align}
\varepsilon<\varepsilon_{\rm crit}:= \max\left(\frac{\varepsilon_\pm}{10},\varepsilon_{\rm diss}\right)
\end{align}
for $m=1$. Since $L_{\rm EM}$ has been determined only for $m=1$, we restrict to the \textit{very} conservative assumption that the spin measurements apply only if Eq.~\eqref{eq:pairprodcrit} is not satisfied for $m>1$, $\varepsilon_{\rm crit}=\varepsilon_\pm/10$. Note, $\varepsilon_\pm$ must be evaluated at the gravitationally saturated cloud mass $M_c=M_c^{\rm sat}$. 

\subsection{Higgs-Abelian model}
When considering a vector field whose mass is generated through a Higgs mechanism involving a coupling to a complex scalar field, there are two main ways this can impact the superradiant growth, saturation, and BH spin-down: (i) at large vector field amplitudes, the cloud undergoes gauged string formation, and (ii) weakly nonlinear vector self-interactions drive an additional radiation channel, dissipating extracted energy of the cloud.

String formation occurs when the amplitude of the superradiance cloud $\max A'^2$ approaches the threshold $A'^2_c=\lambda v^2/g^2$~\cite{East:2022ppo}. In the nonrelativistic limit, the maximum of the vector field amplitude is related to the cloud's total mass by $\max A'^2=\alpha^4M_c/(\pi M)$; relativistic corrections generally correct this down, implying this to be an overestimate for $\max A'^2$ at fixed $M_c$ (see Fig. 7 in Ref.~\cite{Jones:2024fpg}). Therefore, the superradiance cloud spins down the BH, if 
\begin{align}
A'^2_c>A'^2_{\rm string}:=10^2\frac{\alpha^4}{\pi}\frac{M_c^{\rm sat}}{M}.
\label{eq:string_bound}
\end{align}
Here we introduced a conservative factor of $10^2$ following the arguments in Ref.~\cite{Jones:2024fpg}.

Below this critical field amplitude, the superradiance instability proceeds either as in the purely gravitational case, or is slowed down by weakly nonlinear vector radiation. In Ref.~\cite{Jones:2024fpg} (and using the $\alpha$-scaling from Ref.~\cite{Fukuda:2019ewf}), an upper bound for the total energy flux of this additional radiation was obtained for $\alpha\lesssim 0.3$: $\dot{E}^<_{\rm rad}=3 \times 10^{-9} \alpha^6 A'^{-4}_c(M_c/M)^3$. Performing time-domain simulations as in Ref.~\cite{East:2022ppo} for $\alpha\in\{0.4,0.5\}$ (assuming $\chi=0.99$), we obtain flux estimates which suggest a steeper scaling with $\alpha$ for $\alpha>0.3$. Explicitly, in this $\alpha$-range, we find the total massive vector luminosity to be best described by $\dot{E}^>_{\rm rad}=7 \times 10^{-3} \alpha^{18} A'^{-4}_c(M_c/M)^3$. The two flux estimates agree at $\alpha=0.3$ to within $20\%$. We emphasize, these estimates are to be understood with an order-of-magnitude uncertainty. If $\dot{E}^{<,>}_{\rm rad}> \dot{E}_{\rm BH}$ for $\max A'^2<A'^2_{\rm string}$, then the unstable growth of the cloud is quenched before either strings are formed or the instability saturates by spinning down the BH. In the nonrelativistic limit, the flux through the BH horizon is $\dot{E}_{\rm BH}\approx 4\alpha^7\chi M_c/M$. Thus, at the threshold of string formation, which we define to be $\max A'^2=0.1 A'^2_c$, the weakly nonlinear radiation flux reads $\dot{E}_{\rm rad}=3 \times 10^{-11} \alpha^{-2}\pi^2 M_c/M$. From this, the superradiance instability is impacted before string formation occurs for a critical $\alpha$ value,\footnote{The flux $\dot{E}^>_{\rm rad}$ remains orders of magnitude below $\dot{E}_{\rm BH}$ at the string formation threshold.} $\alpha_c \lesssim 0.1$. Therefore, below $\alpha_c$, we must require that at gravitational saturation, the flux ratio remains below unity, $\dot{E}_{\rm rad}/\dot{E}_{\rm BH}|_{\rm sat}< 1$, so there is no impact on the spin down of the BH. This is equivalent to
\begin{align}
A'^2_c>A'^2_{\rm rad}=\sqrt{10}\sqrt{\frac{3\times 10^{-9}}{4\alpha\chi}}\frac{M_c^{\rm sat}}{M},
\end{align}
in the nonrelativistic limit. Note, $A'^2_{\rm rad}\sim \alpha^{1/2}$ for $\alpha\ll 1$. We added the factor of $\sqrt{10}$ to account for the order-of-magnitude uncertainty of our flux estimates above\footnote{The critical $\alpha_c$ is unaffected by this choice.}. Above $\alpha_c$, the condition in Eq.~\eqref{eq:string_bound} applies. Notice, this analysis is valid only for $m=1$; while one may expect string formation to commence at the same threshold even for $m>1$ unstable modes, the vector radiation likely exhibits different scaling with $\alpha$ and overall amplitude. Therefore, we restrict to mapping the spin constraints to bounds on $v\lambda^4$ in the $m=1$ regime.

\subsection{Scalar self-interactions}

The impact of the leading scalar self-interactions on the evolution of the fastest growing superradiant scalar field mode was analyzed in Refs.~\cite{Arvanitaki:2010sy,Baryakhtar:2020gao,Omiya:2022mwv,Omiya:2024xlz}. Self-interactions of this kind lead to a rich phenomenology in the context of superradiance; in this appendix, we focus on those dynamics relevant for BH spin-down induced by the $m=1$ mode only. 

For sufficiently large $f^{-1}$, the self-interactions lead to a pumping of energy from the $m=1$ to the (most unstable) $m=2$ state, while inducing accretion back onto the BH through the $m=0$ decaying mode. This can effectively demote the exponential angular momentum extraction to a linear-in-time process, leading to a potentially long spin-down timescale $\tau_{\rm sd}$. In Ref.~\cite{Baryakhtar:2020gao}, this was estimated to be
\begin{align}
\tau_{\rm sd}=\frac{\sqrt{6}}{m_b}\frac{\Gamma^{322\times \rm BH}_{211\times 211}}{(\omega_I/m_b)^{3/2}(\Gamma^{211\times\infty}_{322\times 322})^{1/2}},
\label{eq:tausd}
\end{align}
where $\Gamma^{322\times \rm BH}_{211\times 211}=4.3\times 10^{-7} \alpha^{11}(M_{\rm pl}/f)^4(1+\sqrt{1-\chi^2})$ and $\Gamma^{211\times\infty}_{322\times 322}=1.1\times 10^{-8}\alpha^8(M_{\rm pl}/f)^4$, with $M_{\rm pl}$ being the Planck mass. If the BH age is smaller than $\tau_{\rm sd}$, the BH is expected to have spun down to the gravitationally saturated state of the $m=1$ mode. We furthermore assume that the expression for the spin-down timescale Eq.~\eqref{eq:tausd} holds only for $\alpha\leq 0.2$ \cite{Baryakhtar:2020gao} and do not set constraints above this value. With this, this timescale is set to $\tau_{\rm sd}=10^5$ years to obtain a conservative upper bound for the self-interaction strength $f^{-1}$. 

\section{Prior and posterior driven constraints}
\label{app:prior_vs_posterior}

\begin{figure*}[htbp]
\includegraphics[width=0.8\textwidth]{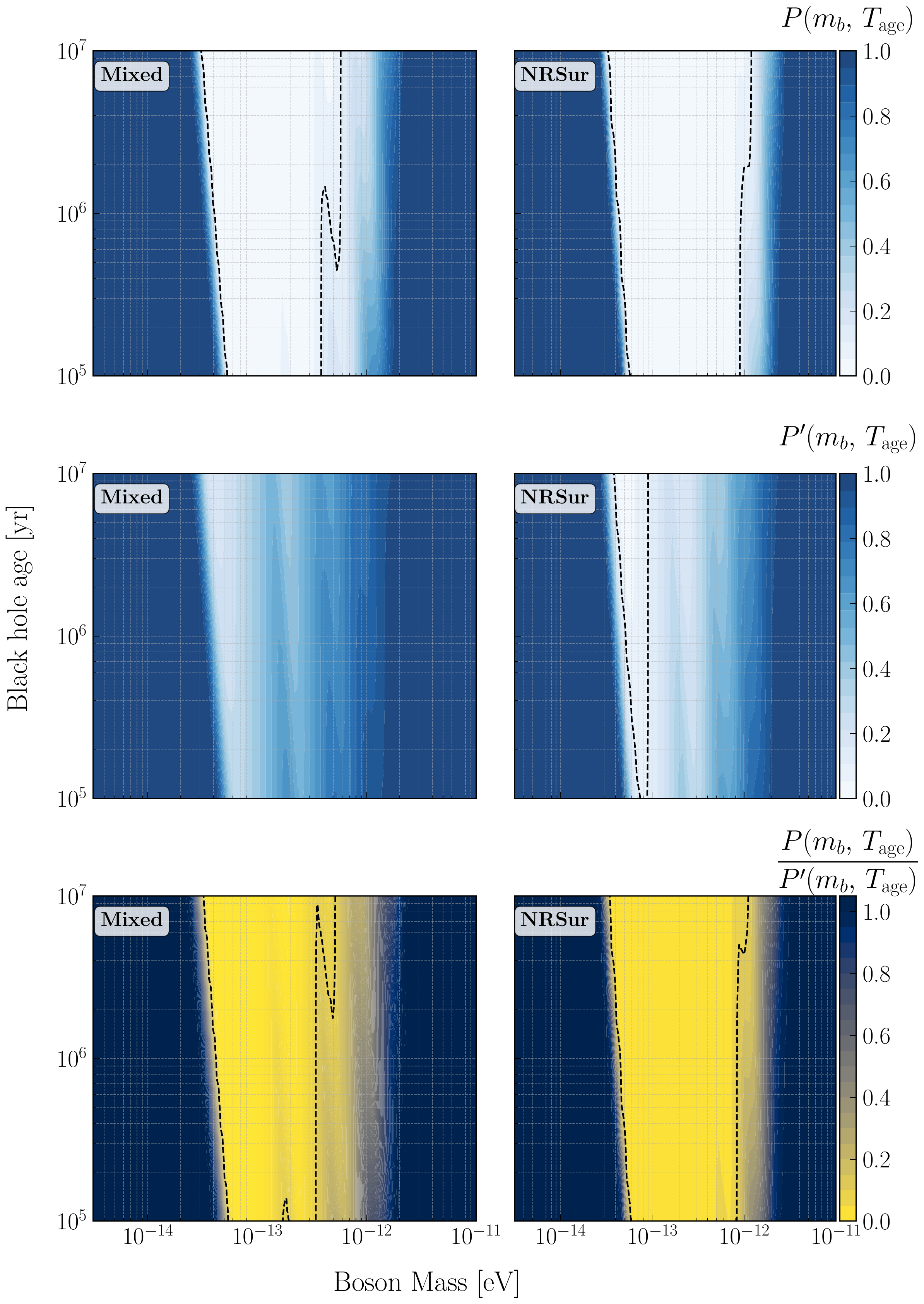}
\caption{Nonexclusion probabilities for a scalar boson mass $m_b$ and BH age $T_{\rm age}$ using GW231123. From top to bottom: $P(\mb, T_{\rm age})$ obtained from posterior distributions, $P'(\mb, T_{\rm age})$ from uniform spin priors, and the ratio $P(\mb, T_{\rm age})/P'(\mb, T_{\rm age})$. Left: Using the primary constituent BH only and the mixed posteriors from five waveform models with equal weight. Right: Using both primary and secondary BHs and the \texttt{NRSur} waveform model. The dashed curves mark the contour with a value of 0.1.}
\label{fig:posterior_prior_comparison}
\end{figure*}

In this section, we discuss the impact of the BH spin prior on the ultralight boson constraints.
In the case that a GW observation is uninformative as to the spin
of a constituent BH, we expect the posterior to be similar to the prior used for this parameter. But as can be seen in Fig.~\ref{fig:chi_max}, a sizable fraction of a uniform distribution in dimensionless spin will be incompatible with an ultralight boson in certain mass ranges.
We want to distinguish the constraints that 
are being driven by the observational measurement of the BH spin from this.
To do this,
we compare the nonexclusion probability calculated from the observation-based posterior $P(\mb, T_{\rm age})$, with that obtained under a uniform spin prior $P'(\mb, T_{\rm age})$. (We note that the calculation of $P'$ still uses the posterior for the BH mass.)
We then identify the 90\% confidence exclusion region where the criterion,
$P(\mb, T_{\rm age}) < 0.1 P'(\mb, T_{\rm age})$, is satisfied. 
Since $P(\mb, T_{\rm age})$ can be interpreted as the fraction of posterior samples that are not excluded for a given boson mass and timescale, and $P'(\mb, T_{\rm age})$ plays the analogous role for the prior, their ratio quantifies how strongly the data disfavors $\mb$. In particular, if the data had no constraining power, we would expect $P(\mb, T_{\rm age}) \approx P'(\mb, T_{\rm age})$. Thus, requiring
${P(\mb, T_{\rm age})}/{P'(\mb, T_{\rm age})} < 0.1$
means that the posterior probability of nonexclusion must drop by at least 90\% compared to the prior expectation. This ensures that the exclusion is driven primarily by the data and corresponds to a 90\% confidence-level exclusion relative to a uniform-spin baseline.

As an example, in Fig.~\ref{fig:posterior_prior_comparison}, we show $P(\mb, T_{\rm age})$, $P'(\mb, T_{\rm age})$, and $P(\mb, T_{\rm age})/P'(\mb, T_{\rm age})$ for the scalar constraints obtained from GW231123.
The left column is obtained from the primary constituent BH with the mixed posteriors from five waveform models (the same as the scenario presented in Fig.~\ref{fig:confidence_plot}), and the right column comes from using both constituent BHs with the posteriors from the \texttt{NRSur} waveform model (which infers a high-spin secondary)~\cite{GW231123,zenodo_GW231123}.  
As demonstrated in the first two rows, although the nonexclusion probabilities are generally higher when adopting a uniform prior (darker in the second row than the first), the prior choice still contributes to the posterior-driven nonexclusion probabilities.
The criterion $P(\mb, T_{\rm age}) < 0.1 P'(\mb, T_{\rm age})$ (enclosed by the dashed curves in the third row) ensures the constraints are posterior-dominated and sets a more restrictive requirement than $P(\mb, T_{\rm age}) < 0.1$ (enclosed by the dashed curves in the first row).

When the secondary constituent BH has a well-constrained high-spin value, including this information in the analysis can improve the constraints. The \texttt{NRSur} waveform model yields $\chi_1 = 0.89^{+0.11}_{-0.20}$ and $\chi_2 = 0.91^{+0.09}_{-0.19}$ for GW231123~\cite{GW231123}. We modify Eq.~\eqref{eq:nonexclusion} to include both BHs as
\begin{widetext}
\begin{equation}
   P(\mb, T_{\rm age}) = \frac{1}{N_{\mathrm{BH}}} \sum_{i=1}^{N_{\mathrm{BH}}} I{\left\{ \chi_1^{(i)} < \chi_{\max}(M_1^{(i)}, \mb, T_{\rm age}) \land \chi_2^{(i)} < \chi_{\max}(M_2^{(i)}, \mb, T_{\rm age}) \right\}},
   \label{eq:nonexclusion2}
\end{equation}
\end{widetext}
where $M_2^{(i)}$ and $\chi_2^{(i)}$ are the secondary BH mass and spin posterior samples, respectively. 
In this case, we calculate $P'(\mb, T_{\rm age})$ by drawing $\chi_1$ and $\chi_2$ from independent uniform distributions over the range of $[0,1)$.
We also note that when using both constituent BHs compared to only the primary, the fraction of the spin prior with two
independent uniform distributions in the
nonexclusion region shrinks noticeably compared to one (middle two panels of Fig.~\ref{fig:posterior_prior_comparison}).
By imposing the condition $P(\mb, T_{\rm age}) < 0.1 P'(\mb, T_{\rm age})$, the constraints are dominated by observation-informed posteriors rather than prior assumptions, even with both constituent BHs. In this case, both the distributions of the two constituent spins and how they are correlated with each other (and the masses) in the posterior will determine the resulting exclusions.  

\section{Constraints from individual waveform models}
\label{app:waveform_comp}

\begin{figure*}[htbp]
\includegraphics[width=\textwidth]{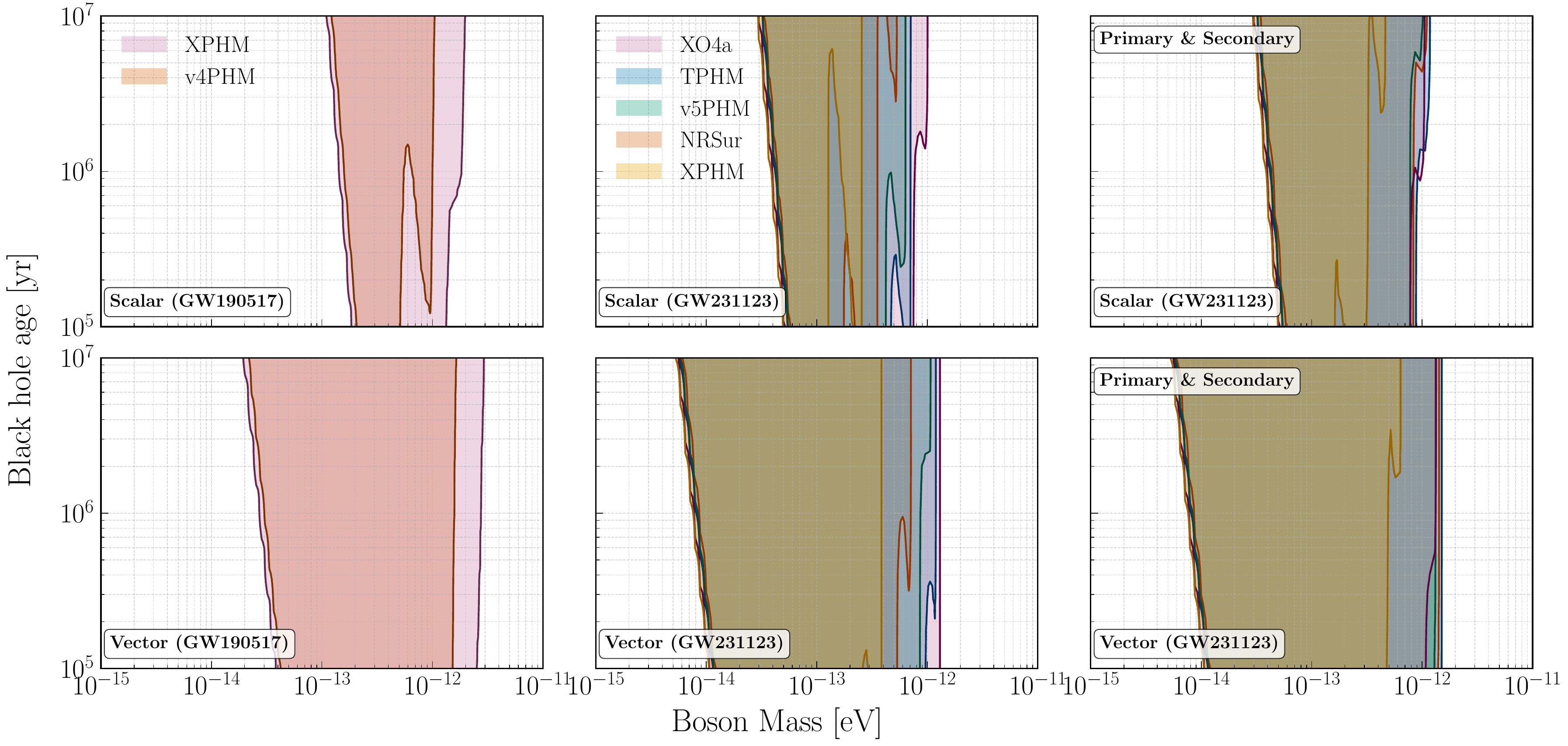}
\caption{Exclusion regions for scalar (top) and vector (bottom) boson masses as a function of the BH age (at confidence levels above 90\%) with individual waveform models. From left to right: GW190517 (primary alone), GW231123 (primary alone), and GW231123 (both constituent BHs).
}
\label{fig:individal_waveform}
\end{figure*}

In this section, we compare the posteriors obtained on BH parameters using different waveform models.
In Fig.~\ref{fig:confidence_plot}, we present results derived from the combined posteriors obtained by equally weighting multiple waveform models for each merger event. 
Although individual waveform models yield slightly different estimates of the binary parameters, we also show the results in Fig.~\ref{fig:individal_waveform} from each model separately to illustrate the level of model-dependent variation.
The complete posterior samples of all waveform models for GW190517 and GW231123 can be found in Refs.~\cite{zenodo_GW190517} and \cite{zenodo_GW231123}, respectively.
For GW190517, all results are derived from the high-spin primary constituent BH alone (left), given that the secondary spin is not well-constrained~\cite{GWTC-2.1}. 
Note that the \texttt{NRSur7dq4} surrogate model is calibrated for a specific region of the binary BH parameter space and is therefore not generally applicable to all systems~\cite{GWTC-2.1,LIGOScientific:2020ibl,Varma:2019csw}. For GW190517, results are available only from the \texttt{XPHM} and \texttt{v4PHM} waveform models~\cite{GWTC-2.1,LIGOScientific:2020ibl,zenodo_GW190517}.
For GW231123, in addition to the results from the primary alone (middle),
we also compute constraints from both constituent BHs for all five waveforms (right).
We note that, as expected, the constraints improve towards higher boson masses when using both BHs, since we are adding information about the lower-mass BH in the binary.
Interestingly, when the secondary is inferred to have a well-constrained high spin, the results that incorporate both BHs also show less variation across waveform models compared to using the primary alone. 
The most distinct results come from \texttt{XPHM}, which infers relatively lower spins for both the primary and secondary ($\chi_1 = 0.79^{+0.21}_{-0.20}$, $\chi_2 = 0.68^{+0.32}_{-0.46}$)~\cite{GW231123}.
For \texttt{XO4a}, which infers a low-spin secondary with $\chi_2 = 0.47^{+0.41}_{-0.47}$, including both BHs does not noticeably improve the constraints compared to using the primary alone.

\end{document}